\documentclass[twocolappendix]{emulateapj}  
\usepackage{color}
\usepackage{epsfig}
\usepackage{subfigure,verbatim}
\usepackage[colorlinks,urlcolor=blue,citecolor=blue,linkcolor=blue]{hyperref}

\renewcommand{\b}[1]{\mbox{\boldmath{$#1$}}} 
\newcommand{\unit}[1]{\nobreak{\mathrm{\;#1}}} 

\newcommand{\be}{\begin{eqnarray}}
\newcommand{\ee}{\end{eqnarray}}
\newcommand{\eq}[1]{Equation~(\ref{eq:#1})}
\newcommand{\fig}[1]{Figure~\ref{fig:#1}}

\def\simless{\mathbin{\lower 3pt\hbox
{$\rlap{\raise 5pt\hbox{$\char'074$}}\mathchar"7218$}}}   
\def\simmore{\mathbin{\lower 3pt\hbox
{$\rlap{\raise 5pt\hbox{$\char'076$}}\mathchar"7218$}}}   

\def\comp{\,c/\omega_{e}}
\def\ompt{\omega_{\rm e}t}
\def\parr{\parallel}
\def\beampar{\Delta p_{b,\parallel}/\gamma_b m_e c}
\def\beamparz{\Delta p_{b0,\parallel}/\gamma_b m_e c}
\newcommand{\ex}[1]{10^{-#1}}
\newcommand{\bmath}[1]{\mbox{\boldmath{$#1$}}}

\begin{document}

\title{Relativistic Pair Beams from T\MakeLowercase{e}V Blazars: A Source of Reprocessed G\MakeLowercase{e}V Emission rather than IGM Heating}

\author{Lorenzo Sironi$^{1,2}$ and Dimitrios Giannios$^{3}$}
\affil{
$^{1}$Harvard-Smithsonian Center for Astrophysics, 60 Garden Street, Cambridge, MA 02138, USA\\
$^{2}$NASA Einstein Postdoctoral Fellow\\
$^{3}$Department of Physics, Purdue University, 525 Northwestern
Avenue, West Lafayette, IN 47907, USA}

\email{E-mail: lsironi@cfa.harvard.edu; dgiannio@purdue.edu.}

\begin{abstract}
The interaction of TeV photons from blazars with the extragalactic background light produces a relativistic beam of electron-positron pairs streaming  through the intergalactic medium (IGM). The fate of the beam energy is uncertain. By means of two- and three-dimensional particle-in-cell simulations, we study the non-linear evolution of dilute ultra-relativistic pair beams propagating through the IGM. We explore a wide range of beam Lorentz factors $\gamma_b\gg 1$ and beam-to-plasma density ratios $\alpha\ll 1$, so that our results can be extrapolated to the extreme parameters of blazar-induced beams ($\gamma_b\sim10^{6}$ and $\alpha\sim10^{{-15}}$, for the most powerful blazars). For cold beams, we show that the oblique instability governs the early stages of evolution, but its exponential growth terminates -- due to self-heating of the beam in the transverse direction -- when only a negligible fraction $\sim(\alpha/\gamma_b)^{1/3}\sim\ex{7}$ of the beam energy has been transferred to the IGM plasma. Further relaxation of the beam proceeds through quasi-longitudinal modes, until the momentum dispersion in the direction of propagation saturates at $\Delta p_{b,\parallel}/\gamma_b m_e c\sim 0.2$. This corresponds to a fraction $\sim 10\%$ of the beam energy being ultimately transferred to the IGM plasma, irrespective of $\gamma_b$ or $\alpha$. If the initial dispersion in beam momentum satisfies $\Delta p_{b0,\parallel}/\gamma_bm_e c\gtrsim 0.2$ (as typically expected for blazar-induced beams),  the fraction of beam energy deposited into the IGM is much smaller than $\sim10\%$. It follows that at least $\sim 90\%$ of the beam energy is still available to power the GeV emission produced by inverse Compton up-scattering of the Cosmic Microwave Background by the beam pairs.
\end{abstract}


\section{Introduction}\label{sec:intro}
With the current generation of $\check{\rm C}$erenkov telescopes, hundreds of TeV sources have been
discovered. By far, the extragalactic TeV sky is dominated by blazars:  jets from galactic centers beaming their emission towards our line of sight. The TeV photons from distant blazars cannot travel cosmological distances, since they interact with the extragalactic  background light (EBL), producing electron-positron pairs.  Studies of the attenuated $\sim100$ GeV$-$TeV light from  distant blazars can therefore provide contraints on the strength of the EBL \citep[e.g.,][]{aharonian_06,abdo_10}.

The produced electron-positron pairs form a relativistic beam moving in the direction of the incident TeV photons. It is usually assumed that the energy of the pair beam is lost via inverse Compton (IC) scattering off the Cosmic Microwave Background (CMB). As a result, the TeV radiation will be reprocessed into the GeV band \citep{neronov_09}. While cooling, the pairs gyrate around the IGM magnetic fields. Depending of the field strength and length scale, the GeV emission may form an extended source, show characteristic delays with respect to the TeV flux, or be strongly suppressed. These effects make combined GeV--TeV studies a useful probe of the IGM fields \citep[e.g.,][]{neronov_10,tavecchio_10,dermer_11,dolag_11,taylor_11,takahashi_12,vovk_12}.

Recently, it has been proposed that the destiny of the blazar-induced beams may be different. As they stream through the  IGM plasma, the electron-positron pairs  are expected to trigger {\it collective} plasma instabilities \citep[as opposed to binary Coulomb collisions, that are negligible, as discussed by][]{miniati_13}. For the parameters relevant to blazar-induced beams (i.e., dilute and ultra-relativistic), the fastest growing mode is the electrostatic oblique instability \citep[e.g.,][]{fainberg_70, bret_10}, whose linear growth rate can exceed the IC cooling rate by several orders of magnitude \citep{brod12a}. {\it Assuming that the instability keeps growing at the linear rate until all the beam energy is deposited into the IGM}, the beam energy loss will be dominated by collective beam-plasma instabilities, rather than IC cooling. In this case, the blazar TeV emission would not be reprocessed down to multi-GeV energies, thus invalidating the IGM field estimates based on the GeV--TeV flux \citep{brod12a,brod13}. In addition, as a result of the beam relaxation, a substantial amount of energy would be deposited into the IGM. This ``volumetric heating''  can have dramatic consequences for the thermal history of the IGM \citep{brod12b,brod12c,brod12d}.

While plasma instabilities could, in principle, be fast enough to thermalize the pair beam, their non-linear stages are far more complicated than what linear dispersion analysis
predicts (e.g., the studies by \citealt{miniati_13} and \citealt{sch12b,sch13} reached opposite conclusions regarding the ultimate fate of blazar-induced beams). 
The nature of the fastest growing instability can change as the beam-plasma system evolves, due to the suppression of temperature-sensitive modes as the beam heats up \citep[e.g.,][]{bret_08}.
Also, beam-plasma instabilities can saturate at very small amplitudes, in particular for the extremely dilute beams produced by TeV blazars \citep[e.g.,][]{thode_75, thode_76}. 

In this work, we use first-principles particle-in-cell (PIC) simulations in two and three dimensions to study the non-linear stages and saturation of the instabilities generated as the blazar-induced pair beams propagate through the IGM. The non-linear effects of the beam-plasma interaction are hard to capture with analytical tools, and they require fully-kinetic simulations. We explore a wide range of beam Lorentz factors $\gamma_b\gg 1$ and beam-to-plasma density ratios $\alpha\ll 1$, so that our results can be extrapolated to the extreme parameters of blazar-induced beams ($\gamma_b\sim10^{6}$ and $\alpha\sim10^{{-15}}$, for the most powerful blazars). We find that, for ultra-relativistic dilute beams that start with a negligible thermal spread, electrostatic beam-plasma instabilities can deposit $\sim10\%$ of the beam energy into the background electrons. However, if the beam is born with a significant momentum dispersion (as expected for blazar-induced beams), the fraction of energy going into IGM heating is much smaller. We conclude that at least $\sim 90\%$ of the beam energy is still available to power the GeV emission produced by IC up-scattering of the CMB. This lends support to the IGM magnetic field estimates that employ the combined GeV--TeV signature of distant blazars.

The paper is organized as follows. In \S\ref{sec:model} we derive the typical parameters of blazar-induced pair beams in the IGM. In \S\ref{sec:setup} we describe the setup of our PIC simulations, whose results are presented in \S\ref{sec:results}. In particular, in \S\ref{sec:time} we focus on one representative choice of beam parameters (in the regime of dilute ultra-relativistic cold beams) and we describe the complete evolution of the beam-plasma unstable system, from the early exponential phase up to the non-linear stages. In \S\ref{sec:hot}, we discuss the dependence of our findings on the beam Lorentz factor, the beam-to-plasma density contrast and the beam temperature. For the convenience of readers uninterested in the kinetic details of  beam-plasma instabilities, in \S\ref{sec:blazar} we summarize our results in application to blazar-induced beams. Finally,  in \S\ref{sec:astro} we assess the implications of our findings for the thermal history of the IGM and the detection of reprocessed GeV emission from powerful TeV blazars.

\section{Physical Parameters of Blazar-Driven Beams}\label{sec:model}
In this section, we summarize the physical parameters of blazar-induced beams, including the density contrast to the IGM, the beam Lorentz factor and   velocity spread. We present order-of-magnitude estimates, and we refer to \citet{sch12a} and  \citet{miniati_13} for a more detailed analysis of the beam distribution function, and its dependence on the spectrum of the EBL and of the blazar TeV emission.

Blazar photons of energy $E_\gamma\sim 10\unit{TeV}$ travel a distance  of
$D_{\gamma\gamma}\simeq 80\, K_{\rm{EBL}}\,(E_\gamma/10\,\rm TeV)^{-1}\unit{Mpc}$ before they interact with the EBL and produce electron-positron pairs \citep{neronov_09}. Here, $K_{\rm EBL}\sim 1$ accounts for uncertainties in the intensity of the EBL, with models predicting
$0.3\simless K_{\rm EBL}\simless 3$ for $0.1\, {\rm TeV}\simless E_\gamma\simless 10\, \rm TeV$ \citep[e.g.,][]{aharonian_01}.\footnote{Throughout the paper, we neglect the dependence on cosmological redshift. Strictly speaking, our results apply to
$z\sim 0$, but they can be easily generalized to arbitrary redshifts, provided that one makes additional assumptions about the redshift evolution
of the EBL and of the blazar luminosity function.} Each particle moves along the direction of the incident TeV photon, and it carries about half of the photon energy, so the beam Lorentz factor  is $\gamma_b\simeq10^7 (E_\gamma/10\,\rm TeV)$. Assuming that plasma instabilities in the IGM do not appreciably affect the beam propagation (an assumption that is correct {\it a posteriori}, as we demonstrate in this work), the pairs  travel a distance of
$d_{\rm IC}\simeq100\, (E_\gamma/10\,\rm TeV)^{-1}\unit{kpc}$ before cooling by  IC scattering off the CMB (leading to many photons of energy $\sim 100\,(E_\gamma/10\,\rm {TeV})^2\unit{GeV}$ per original TeV photon). 

For a powerful blazar with TeV isotropic equivalent luminosity $L_\gamma\simeq10^{45}L_{\gamma,45}\unit{erg\, s^{-1}}$ \citep[e.g.,][]{ghisellini_10}, the number density $n_b$ of the beam pairs is set by the balance of the pair production rate with the energy loss rate (dominated by IC cooling), which gives 
\be
\!\!\!\!\!n_b\!\sim\! 2 \frac{L_{\gamma}/E_{\gamma}}{4 \pi D_{\gamma\gamma}^2 c }\frac{d_{\rm IC}}{D_{\gamma\gamma}}\!\simeq\! 10^{-23}  K_{\rm EBL}^{-3}L_{\gamma,45}\!\! \left(\!\frac{E_\gamma}{10\unit{TeV}}\!\right)\!\!\unit{cm^{-3}}
\ee
where the factor of $d_{\rm IC}/D_{\gamma\gamma}\ll 1$ accounts for the rapid energy loss of the pairs due to IC.\footnote{If plasma instabilities were to dominate the energy loss of the beam, $d_{\rm IC}$ should be replaced by the beam thermalization length.} If the number density in the IGM  is $n_{\rm IGM}\sim 10^{-7}$ cm$^{-3}$ (but it may be a factor of several smaller in cosmological voids, which dominate the cosmic space at $z\sim 0$), the density ratio between the streaming pairs and the background plasma is
\be
\!\!\alpha\!\equiv \!\frac{n_b}{n_{\rm IGM}}\!\simeq 10^{-16}\! K_{\rm EBL}^{-3}L_{\gamma,45} \!\! \left(\!\frac{E_\gamma}{10\unit{TeV}}\!\right)\!\!\left(\!\frac{n_{\rm IGM}}{10^{-7}\unit{cm^{-3}}}\!\right)^{-1}
\ee
The density contrast $\alpha$ and the beam Lorentz factor $\gamma_b$ are the two crucial parameters determining the plasma physics of the beam-IGM interaction. For blazar-induced beams, we expect that they should vary in the range $\alpha\sim 10^{-18}-10^{-15}$ and $\gamma_b\sim 10^6-10^7$, respectively. 

As discussed by \citet{brod12a}, collective beam-plasma effects can be relevant only if many beam pairs are present within  a sphere of radius equal to the wavelength of the most unstable mode. As we show below, the scale of the fastest growing modes is $2\pi \,c/\omega_e$, where $c/\omega_e=\sqrt{m_e c^2/4 \pi e^2 n_e}\simeq1.5\times10^9\, (n_e/10^{-7}\!\unit{cm^{-3}})^{-1/2}\unit{cm}$ is the plasma skin depth of the IGM electrons (with number density $n_e=n_{\rm IGM}/2$). A sphere of skin-depth radius contains
$(2\pi c/\omega_e)^3n_b\sim 10^7\, (\alpha/\ex{16})(n_e/10^{-7}{\rm cm}^{-3})^{-1/2}$ beam particles. For $\alpha\sim 10^{-18}-10^{-15}$, we find that collective phenomena always play a role in the evolution of blazar-induced beams.

Another important parameter is the dispersion in beam momentum at birth. Since the pair creation cross section peaks slightly above the threshold energy, the pairs are born moderately warm (with a comoving temperature of $k_{\rm B} T_b\simeq0.5\, m_e c^2$). Moreover, since the EBL and the blazar TeV spectra are broad, the beam energy distribution will extend over a wide range of Lorentz factors, as discussed by \citet{miniati_13}. In \S\ref{sec:hothot}, we explore the role of thermal effects on the non-linear evolution of blazar-induced beams.

\section{Simulation Setup}\label{sec:setup}
We investigate blazar-driven plasma instabilities in the IGM by means of fully-kinetic PIC simulations. We employ the three-dimensional (3D) electromagnetic PIC code TRISTAN-MP \citep{spitkovsky_05}, which is a parallel version of the publicly available code TRISTAN \citep{buneman_93}, that was optimized for handling ultra-relativistic flows (see, e.g., \citealt{sironi_spitkovsky_11a} and \citealt{sironi_13} for studies of ultra-relativistic collisionless shocks using TRISTAN-MP). We initialize a relativistic dilute pair beam that propagates along  $+\bmath{\hat{x}}$ through an unmagnetized electron-proton plasma (with the realistic mass ratio $m_p/m_e=1836$). The simulations are performed in the frame of the background plasma, i.e., of the IGM. No background magnetic field is assumed, so the electric and magnetic fields generated by beam-plasma instabilities will grow from noise.

To follow the beam-plasma evolution to longer times with fixed computational resources, we mainly utilize 2D computational domains in the $xy$ plane. In \S\ref{sec:time} we compare 2D and 3D runs, and we show that 2D simulations can capture most of the relevant 3D physics. In the case of 2D simulations with the beam lying in the simulation plane, only the in-plane components of the
velocity, current and electric field, and only the out-of-plane component of the magnetic field are present. The simulation box is periodic in all directions. By choosing a periodic domain, we simulate the bulk of the beam-plasma system, rather than the ``head'' of the pair beam. 

The background plasma consists of cold electrons and protons, with initial electron temperature $k_{\rm B} T_e/m_e c^2\simeq10^{-8}$. We have tested that   higher temperatures of the background electrons do not significant change the development of the relevant instabilities, as long as the electron temperature is non-relativistic, in agreement with \citet{bret_05}. The background protons are allowed to move, but we obtain similar results when the protons are treated as  a static charge-neutralizing background.  

The beam consists of electron-positron pairs propagating with Lorentz factor $\gamma_b$ along the $+\bmath{\hat{x}}$ direction. To our knowledge, our PIC simulations are the first to address the evolution of an electron-positron beam. All of the previous studies have focused on the case of an electron beam propagating through an electron-proton plasma, with the background electrons moving opposite to the beam to compensate for the beam current \citep[e.g.,][]{dieckmann_06,bret_08,kong_09}. Any instability triggered by the relative drift between the background electrons and protons (e.g., the \citet{buneman_58} instability) will then be absent in our setup, where the pair beam carries no net current.

The beam-to-plasma density ratio $\alpha$ and the beam Lorentz factor $\gamma_b$ expected for blazar-induced pairs streaming through the IGM (see \S\ref{sec:model}) cannot be directly studied with PIC simulations. Yet, by performing dedicated experiments with a broad range of $\alpha$ and $\gamma_b$ (in the regime $\alpha\ll1$ and $\gamma_b\gg1$ of ultra-relativistic dilute beams), we can extrapolate the relevant physics to the extreme parameters expected in the IGM. We vary the beam Lorentz factor from $\gamma_b=3$ up to $\gamma_b=1000$, and the density contrast from $\alpha=10^{-1}$ down to $\alpha=10^{-3}$.\footnote{Beams with more extreme parameters (in particular, with $\alpha\lesssim \ex{3}$) will take longer to evolve, and at that point the fact that explicit PIC codes do not conserve energy to machine precision (see below) can be a limitation for the reliability of our results.}  For numerical convenience, the density ratio between the beam and the plasma is established by initializing the same number of beam and plasma computational particles,  with the beam particles having a weight $\alpha$. We have tested that, by choosing a different weight (yet, keeping the same physical density contrast), our results do not change. In addition to studying  the dependence on $\gamma_b$ and $\alpha$, we also compare the evolution of cold beams (with comoving temperature at initialization $k_{\rm B} T_{b}/m_ec^2\simeq10^{-4}$) with the case of warm beams, up to the limit of mildly relativistic thermal spreads $k_{\rm B} T_{b}/m_ec^2\sim1$ most relevant for blazar-induced beams.

The results presented below have been extensively tested for convergence. We typically employ 50  particles per computational cell for the background plasma (25 electrons and 25 protons), and the same number for the beam particles (if each carries a weight $\alpha$). However, we have tested that our results are the same when using up to 256  particles per cell, for both the beam and the plasma. We resolve the skin depth $c/\omega_e$ of the background electrons  with 8 computational cells, but we have tested that our results do not change when using $12$ or $16$ cells per skin depth.\footnote{The speed of light in the simulations is 0.45 cells/timestep, so that the temporal resolution is $\delta t=0.05625\,\omega_e^{-1}$.} In 2D runs, the simulation plane is typically a square with $1024$ cells ($\sim125\,c/\omega_e$) on each side, but we have checked that our results do not substantially change when employing a larger box, that is $500\,c/\omega_e$ long (in the direction of beam propagation) and $250\,c/\omega_e$ wide. In 3D we employ a box with $512$ cells ($\sim67.5\,c/\omega_e$) in the transverse direction and $1024$ cells ($\sim125\,c/\omega_e$) along the longitudinal direction.\footnote{Hereafter, ``longitudinal'' and ``transverse" will be relative to the beam direction of motion.} To capture the linear and non-linear stages of the beam-plasma evolution, we follow the system up to unprecedentedly long times, in 2D up to $\omega_e t\sim 10^5 $, or equivalently $\sim 1.75\times10^6$ timesteps, and in 3D up to $\omega_e t\sim 4\times10^4 $, or $\sim 7\times10^5$ timesteps.

The number of beam particles is kept constant during the evolution of the beam-plasma system, since the photon-photon interactions that would introduce fresh electron-positron pairs are extremely rare on the timescales covered by our simulations.  Also, we neglect IC cooling of the beam pairs, since it is irrelevant over the timespan of our runs. This implies that the total energy in our periodic beam-plasma system should be constant over time. However,  explicit PIC codes do not conserve energy to machine  precision. We track the energy conservation in our runs, and we find that at  late times it is still better than $1\%$. This makes our estimates of the amount of beam energy transferred to the plasma electrons (of order $\sim 10\%$) extremely robust, for the beam parameters explored in this work. 

Finally, we remark that in all PIC codes a numerical heating instability arises when cold relativistic plasma propagates for large distances over the numerical grid \citep{dieckmann_06b}. Since the numerical speed of light on the grid is smaller than the correct value at large wavenumbers, ultra-relativistic particles will emit numerical $\check{\rm C}$erenkov radiation. This might artificially slow down the beam, even in the absence of physical beam-plasma instabilities. We have assessed that the results reported below arise from a physical instability (as opposed to the numerical $\check{\rm C}$erenkov mode), by comparing our beam-plasma simulations with the artificial case of a beam that propagates through the grid in the absence of any background plasma. The beam evolution in the two cases is dramatically different, which provides further confirmation that the beam energy loss that we discuss below arises from the physical interaction of the beam with the background plasma, rather than from the numerical $\check{\rm C}$erenkov instability.

\section{Results}\label{sec:results}
In this section, we explore the linear and non-linear evolution of ultra-relativistic dilute pair beams by means of 2D and 3D PIC simulations. In \S\ref{sec:time}, we describe the different stages of evolution of the beam-plasma system, for a representative choice of beam parameters in the regime of ultra-relativistic dilute beams ($\gamma_b=300$, $\alpha=\ex{2}$ and negligible beam thermal spread at initialization). In \S\ref{sec:hot}, we investigate the dependence of our results -- in particular, of the fraction of beam energy transferred to the background electrons -- on the beam-to-plasma density contrast, the beam Lorentz factor and the beam temperature at birth. The reader that is not interested in the kinetic details of the beam-plasma interaction might proceed to \S\ref{sec:blazar}, where we extrapolate the findings of our PIC simulations to the extreme parameters of blazar-induced beams.

\begin{figure*}[tbp]
\begin{center}
\includegraphics[width=1.05\textwidth]{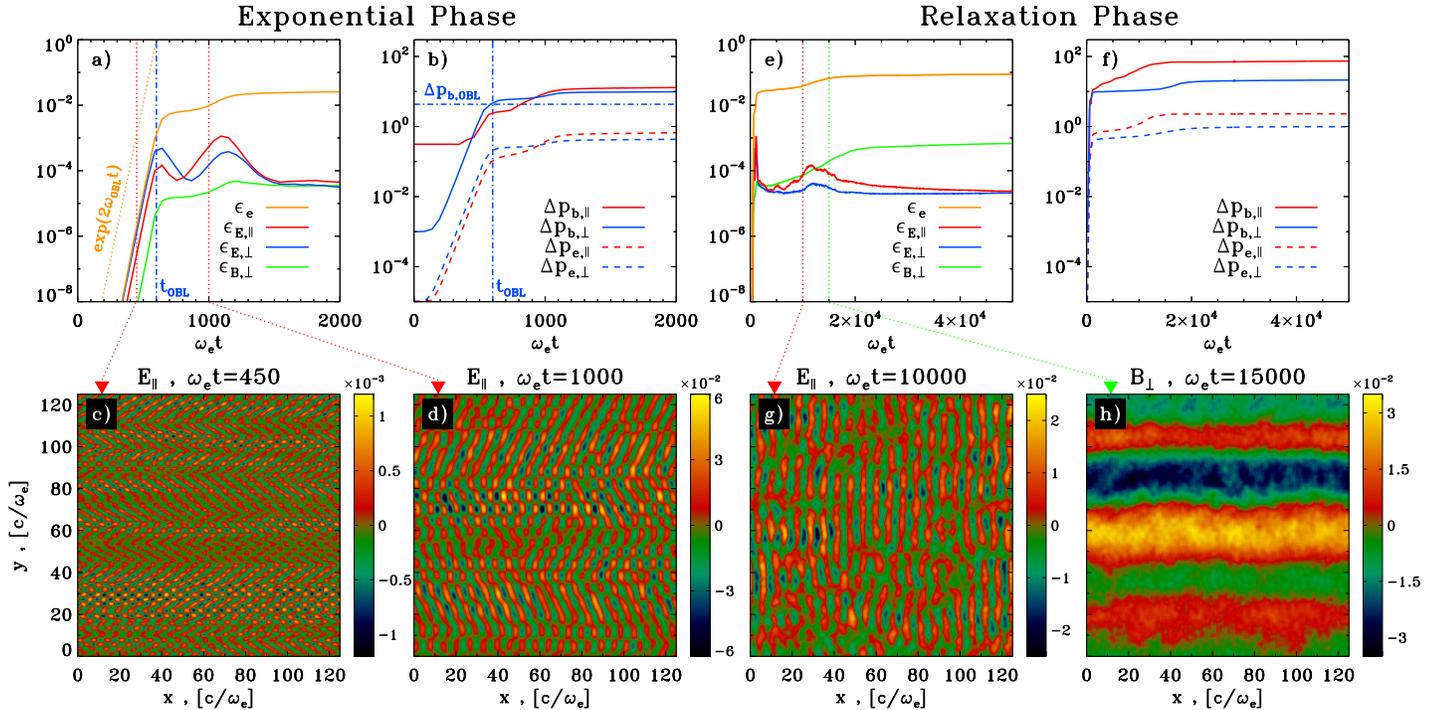}
\caption{Temporal evolution of the beam-plasma interaction, from the 2D simulation of a cold beam with $\gamma_b=300$ and $\alpha=\ex{2}$. We follow the evolution of the system through the exponential phase of the oblique mode (panels (a)-(d)) until the relaxation stage (panels (e)-(h)). Panels (a) and (e): fraction of beam kinetic energy transferred to the plasma electrons (orange), to the longitudinal and transverse electric fields (red and blue, respectively) and to the transverse magnetic fields (green). In panel (a), the dotted orange line shows the growth rate of the oblique mode expected from linear dispersion analysis. Panels (b) and (f): temporal evolution of the momentum dispersion of the beam (solid) and plasma (dashed) electrons, along the beam (red) or transverse to the beam (blue). The momenta are in units of $m_e c$. Panels (c),(d) and (g): 2D plot of the longitudinal electric field $E_\parallel$, in units of $\sqrt{8 \pi \gamma_b n_b m_e c^2}$. The electric field is shown at three different stages of evolution, as marked by the red arrows at the bottom of panels (a) and (e). Panel (h): 2D plot of the transverse magnetic field $B_\perp$, in units of  $\sqrt{8 \pi \gamma_b n_b m_e c^2}$, at the time marked by the green arrow  at the bottom of panel (e).}
\label{fig:timeevol}
\end{center}
\end{figure*}

\begin{figure*}[tbp]
\begin{center}
\includegraphics[width=1.03\textwidth]{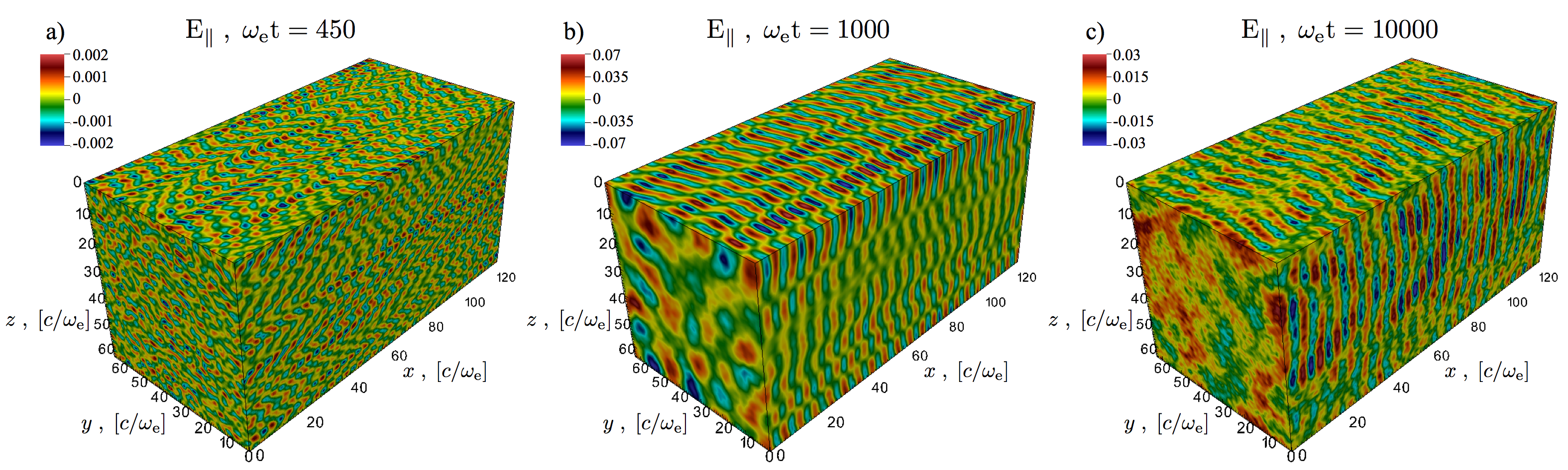}
\caption{3D structure of the longitudinal electric field $E_{\parallel}$, from the 3D simulation of a cold beam with $\gamma_b=300$ and $\alpha=\ex{2}$. The electric field is in units of $\sqrt{8 \pi \gamma_b n_b m_e c^2}$. The three snapshots are taken at the same times of panels (c), (d) and (g) in \fig{timeevol}, and they show that the 3D physics of the relevant electrostatic beam-plasma instabilities can be correctly captured by our 2D runs. In the 3D simulations, the temporal evolution of the fraction of beam kinetic energy transferred to the background electrons and to the electromagnetic fields (not shown here), as well as the time evolution of the beam and plasma momentum dispersions (still not shown here), closely follow the 2D results presented in \fig{timeevol}(a),(b),(e) and (f). The equivalence of 2D and 3D results for cold beams is indeed expected on analytical grounds \citep[e.g.,][]{bret_10}.}
\label{fig:3d}
\end{center}
\end{figure*}

\subsection{The Linear and Non-Linear Evolution of Ultra-Relativistic Dilute Cold Beams}\label{sec:time}
In this section, we follow the evolution of a cold beam with $\gamma_b=300$ and $\alpha=\ex{2}$. We start with the analysis of the linear phase, and then we investigate the non-linear relaxation. We find that the exponential phase of the oblique mode (which is the fastest growing instability for dilute ultra-relativistic beams) terminates due to self-heating of the beam in the direction transverse to the beam motion. At the end of the oblique phase, only a minor fraction $\sim (\alpha/\gamma_b)^{1/3}$ of the beam energy has been deposited into the background electrons. Further evolution of the beam is governed by quasi-longitudinal modes, which operate on a timescale that is much longer (at least two orders of magnitude) than the oblique growth. At the end of the quasi-longitudinal phase, the dispersion of beam momentum in the longitudinal direction saturates at $\Delta p_{b,\parallel}/\gamma_b m_e c\sim0.2$, which corresponds to a fraction $\sim 10\%$ of  beam energy transferred to the plasma.

\subsubsection{The Oblique Exponential Phase}\label{sec:oblique}
The evolution of the beam-plasma system during the oblique phase is presented in panels (a)-(d) of \fig{timeevol}. The  beam is set up with a small thermal spread ($k_{\rm B} T_b/m_e c^2\simeq10^{-6}$), so that the oblique instability initially proceeds in the  {\it reactive} regime, i.e., all the beam particles are in resonance with each harmonic of the packet of unstable modes, and the instability is the strongest. This is opposed to the { \it kinetic} regime, in which the beam velocity spread is considerable. Here, a number of unstable modes with a broad spectrum in phase velocity  will be excited, with only a small number of beam particles being in resonance with each mode. This results in a slower growth, as compared to the reactive regime.

In \fig{timeevol}, we confirm that the oblique instability is the fastest growing mode for ultra-relativistic dilute beams. The reactive phase of the  instability governs the evolution of the system for $t\lesssim t_{\rm OBL}\simeq 600\,\omega_e^{-1}$, where $t_{\rm OBL}$ is marked as a dash-dotted vertical blue line in panels (a) and (b). Here, $\omega_e=\sqrt{4\pi e^2 n_e/m_e}$ is the plasma frequency of the background electrons. The fastest mode grows on a scale $\sim 2\pi c/\omega_e$, where $ c/\omega_e$ is the electron skin depth, and its wavevector  is inclined at $\sim 45^\circ$ with respect to the beam propagation. This is apparent in the 2D structure of the longitudinal electric field in \fig{timeevol}(c), as well as in the 2D plots of the transverse electric and magnetic fields (not shown).\footnote{We remind that, for 2D runs with the beam lying in the simulation plane, only the in-plane components of the electric field, and the out-of-plane component of the magnetic field are present.} The oblique mode is also captured in 3D simulations, as shown in the 3D structure of the longitudinal electric field of \fig{3d}(a).

The growth rate of the  oblique instability in the reactive regime is \citep[e.g.,][]{fainberg_70}
\be\label{eq:omr}
\omega_{\rm OBL}({\b k})=\frac{\sqrt{3}}{2^{4/3}}\left(\frac{2 \alpha}{\gamma_b}\right)^{1/3}\left(\frac{k_\perp^2}{k^2}+\frac{k_\parallel^2}{\gamma_b^2k^2}\right)^{1/3}\omega_e~~,
\ee
where the different dependence on $k_\perp$ and $k_\parallel$ is related to the fact that for relativistic beams the transverse inertia is much smaller than the longitudinal inertia (by  a factor of $\gamma_b^2$), so that  the modes transverse to the beam are the easiest to be excited (for an intuitive physical description, see \citealt{nakar_11}).\footnote{In \eq{omr}, the factor of $2$ that multiplies $\alpha$ is related to our definition of $\alpha=n_b/n_{\rm IGM}=n_b/2\,n_e$.} From the pattern in \fig{timeevol}(c), we infer $k_\perp\sim k_{\parallel}\sim k/\sqrt{2}$, so the growth rate of the fastest growing oblique mode will be
\be\label{eq:ommax}
\omega_{\rm OBL}\sim \frac{\sqrt{3}}{2^{4/3}}\left(\frac{\alpha}{\gamma_b}\right)^{1/3}\!\!\omega_e\equiv \delta_{\rm OBL} \,\omega_e~~,
\ee
which nicely agrees with our results. In fact, in \fig{timeevol}(a) we show that the fraction of beam kinetic energy deposited into the background electrons (orange line), into the longitudinal (red) and transverse (blue) electric fields, and into the transverse magnetic field (green) all grow at the rate predicted by \eq{ommax} for $t\lesssim 600\,\omega_e^{-1}$ (dotted orange line in \fig{timeevol}(a)). 

The oblique mode is quasi-electrostatic, i.e., roughly $ \bmath{k}\parallel \bmath{E_k}$ \citep[e.g.,][]{bret_10}. Since the angle between the wavevector and the beam is $\gtrsim 45^\circ$, in agreement with analytical expectations \citep[e.g.,][]{bret_10}, it follows that the electric field component tranverse to the beam is slightly larger than the longitudinal component, i.e., $k_\perp\gtrsim k_\parallel$ implies that $E_\perp\gtrsim E_\parallel$. This explains the small difference between the red and the blue lines in \fig{timeevol}(a). Also, since the mode is quasi-electrostatic, the magnetic component will be sub-dominant relative to the electric fields. In agreement with the analytical considerations of \citet{pelletier_10}, we find that $B_\perp\sim2\, \delta_{\rm OBL} \,E_{\perp}$. Given that $\delta_{\rm OBL}\ll1$ for ultra-relativistic dilute beams, it follows that $B_\perp\ll \,E_{\perp}$.

For electrostatic modes in the linear phase, it is expected that the amount of kinetic energy lost by the beam should be equally distributed between electric fields and plasma heating \citep[e.g.,][]{thode_76}. This explains why the energy $\epsilon_e$ of the background electrons in the exponential phase (orange line in \fig{timeevol}(a) at $t\lesssim t_{\rm OBL} \simeq 600\,\omega_e^{-1}$) is comparable to the energy in electric fields ($\epsilon_{E,\parallel}$ and $\epsilon_{E,\perp}$, respectively red and blue lines in \fig{timeevol}(a)). We have verified that the fraction of beam energy transferred to the background protons is negligible as compared to the plasma electrons, so the evolution of the protons will be ignored hereafter.

Since $E_\perp\sim E_\parallel$ during the exponential phase of the oblique mode, both the plasma and the beam are heated quasi-isotropically, so that the momentum spreads in the longitudinal and transverse directions are nearly identical (compare the red and blue lines for $t\lesssim t_{\rm OBL}$ in \fig{timeevol}(b); dashed lines refer to the plasma, solid lines to the beam). The transverse momentum spread of the beam can be related to the transverse electric field $E_{\perp}$ via the Lorentz force
\be\label{eq:lorentz}
\frac{\Delta p_{b,\perp}}{\Delta t}\sim \delta_{\rm OBL}\,\omega_e \Delta p_{b,\perp} \sim e E_{\perp}~~,
\ee
where we have assumed that the characteristic timescale is set by the oblique growth rate (i.e., $\Delta t^{-1}\sim \delta_{\rm OBL}\,\omega_e$) and that the magnetic force is negligible compared to the electric force (in fact, $B_{\perp}/E_{\perp}\sim 2\, \delta_{\rm OBL}\ll1$).

Since the oblique mode is heating up the beam in the transverse direction (solid blue line in \fig{timeevol}(b)), the exponential growth at the reactive rate $\omega_{\rm OBL}$ will necessarily terminate, when the assumption of a cold beam required by the reactive approximation becomes invalid. It is well known that the system will transition from the reactive phase to the kinetic phase when the beam velocity dispersion $\bmath{\Delta v_b}$ reaches \citep[e.g.,][]{fainberg_70,bret_10}
\be\label{eq:sat}
|\bmath{k}\cdot\bmath{\Delta v_b}|\sim \omega_{\rm OBL}~~,
\ee
namely when the beam, due to its velocity spread, can move across one wavelength of the most unstable mode during the growth time of the reactive instability. In this case, most of the beam particles will lose resonance with the unstable mode, and the instability will transition from the reactive to the kinetic regime. For ultra-relativistic beams with isotropic momentum dispersions (in fact, \fig{timeevol}(b) shows that $\Delta p_{b,\perp}\sim \Delta p_{b,\parallel}$ during the oblique reactive phase), the transverse velocity spread $\Delta v_{b,\perp}/c\sim \Delta p_{b,\perp}/\gamma_b m_e c$ is much larger than the longitudinal spread $\Delta v_{b,\parallel}/c\sim (\Delta v_{b,\perp}/c)^2+\Delta p_{b,\parallel}/\gamma_b^3m_e c$. Since $k_\perp\sim k_\parallel\sim \omega_e/c$, \eq{sat} above reduces to
\be\label{eq:thresh}
\Delta p_{b,\rm OBL}\sim \delta_{\rm OBL} \gamma_b\,m_e c~~,
\ee
where $\Delta p_{b,\rm OBL}$ is the expected transverse dispersion in beam momentum at the end of the reactive oblique phase. The threshold in momentum dispersion $\Delta p_{b,\rm OBL}$ can also be recast as a limit in beam temperature \citep[e.g.,][]{bret_10b}. We confirm that the reactive phase of the oblique mode terminates at $t\sim t_{\rm OBL}\sim 600\,\omega_e^{-1}$, when the beam transverse momentum reaches the threshold $\Delta p_{b,\rm OBL}$ in \eq{thresh} (which is shown as a horizontal dash-dotted blue line in \fig{timeevol}(b)). 

We can now derive the expected fraction of beam kinetic energy transferred to the plasma electrons and to the electromagnetic fields at the end of the oblique reactive phase. By setting $\Delta p_{b,\perp}=\Delta p_{b,\rm OBL}$ in \eq{lorentz}, we find that the fraction of beam energy converted into transverse electric fields at $t\sim t_{\rm OBL}$ is 
\be\label{eq:eperp}
\epsilon_{{E,\perp}}\equiv\frac{E_\perp^2}{8 \pi \gamma_b n_b m_e c^2}\sim\frac{\sqrt{27}}{32}\delta_{\rm OBL}~~.
\ee
Since the oblique mode is quasi-electrostatic (i.e., $E_\perp\sim E_{\parallel}$), it follows that $\epsilon_{{ E,\perp}}\sim \epsilon_{{ E,\parallel}}$. Moreover, for electrostatic modes, the fraction $\epsilon_e$ of the beam kinetic energy converted into plasma heating is comparable to the energy in electric fields \citep[e.g.,][]{thode_76}, so $\epsilon_e\sim\epsilon_{{ E,\perp}}\sim \epsilon_{{ E,\parallel}}$. Finally, since $B_\perp\sim 2 \,\delta_{\rm OBL}\, E_{\perp}$, the magnetic energy fraction will be $\epsilon_{B,\perp}\sim \sqrt{27}\,\delta_{\rm OBL}^3/8 $. We have extensively verified that the expected scalings of the  efficiency parameters $\epsilon_e,\,\epsilon_{{ E,\perp}},\, \epsilon_{{ E,\parallel}}$ and $\epsilon_{{ B,\perp}}$ with respect to $\delta_{\rm OBL}$ are in agreement with the results of our simulations, across the whole range of beam Lorentz factors and density contrasts we have explored (see the various curves in \fig{timeevol}(a) at $t\sim t_{\rm OBL}$, and also \S\ref{sec:cold}).

For $t\gtrsim t_{\rm OBL}$, the evolution of the oblique mode will proceed in the kinetic (rather than reactive) regime. The kinetic oblique mode is indeed resposible for the peak in the electric field energy observed at $\omega_e t\sim 1000$ in \fig{timeevol}(a) (red and blue lines), which produces a moderate increase in the fraction of beam energy transferred to the background electrons (orange line in  \fig{timeevol}(a) at $\omega_e t\sim 1000$). In this phase, the  2D structure of the longitudinal electric field in \fig{timeevol}(d) shows that the wavevector of the kinetic oblique mode is oriented at $\sim20^\circ$ relative to the beam propagation (as compared to the $\sim 45^\circ$ angle observed during the reactive phase, see \fig{timeevol}(c)). A similar pattern is shown in the 3D plot of \fig{3d}(b). As expected, the increase in the transverse momentum dispersion has suppressed the modes having $k_\perp\gg k_\parallel$, which are most sensitive to transverse temperature effects \citep[e.g.,][]{bret_10}.

For a beam with initial transverse velocity dispersion $\Delta v_{0,\perp}$, the growth rate of the kinetic oblique mode for $k_\perp\lesssim \omega_e/c$ is \citep[e.g.,][]{breizman_ryutov_71}
\be\label{eq:kin}
\omega_k\sim \left(\frac{c}{\Delta v_{0,\perp}}\right)^2 \frac{\alpha}{\gamma_b}\,\omega_e\equiv \delta_{k} \,\omega_e~~.
\ee
At the end of the reactive oblique phase, \eq{thresh} prescribes that $\Delta v_{0,\perp}/c=\Delta p_{b,\rm OBL}/\gamma_b m_e c\sim \delta_{\rm OBL}$, so that the growth rate of the kinetic oblique mode will be $\omega_k\propto \delta_{\rm OBL}\, \omega_e$, i.e., it will have the same scalings with $\alpha$ and $\gamma_b$ as the reactive oblique mode (here, we have neglected factors of order unity).

We have explicitly verified that the peak in electric fields at $\omega_e t\sim 1000$ is due to the kinetic oblique mode, by performing a dedicated simulation in which at $\omega_e t\sim 800$ (i.e., shortly after the end of the reactive stage) we reset by hand the electromagnetic fields and the plasma temperature to their initial values (i.e., no seed fields and $k_{\rm B} T_e/m_e c^2\simeq10^{-8}$), yet we retain the beam momentum distribution that results self-consistently from the reactive oblique phase. In this setup, we find that the fastest growing mode has the same 2D pattern as in \fig{timeevol}(d) and its growth rate scales as $\propto \delta_{\rm OBL}\, \omega_e$. This confirms that the peak in electric fields at $\omega_e t\sim 1000$ (\fig{timeevol}(a)) is indeed associated to  the kinetic oblique instability.

The growth of the kinetic oblique mode terminates due to self-heating of the beam, in analogy to the reactive oblique phase. In particular, the growth in  the kinetic oblique phase cannot be sustained beyond the point where, due to the self-excited electric fields, the beam momentum dispersion in the transverse direction exceeds the initial value $\gamma_b\Delta v_{0,\perp}m_e$. At this point, the expression for the growth rate in \eq{kin} becomes clearly invalid. This happens when
\be
e E_{\perp}\sim \omega_k \gamma_b\, \Delta v_{0,\perp} m_e  \sim \delta_{\rm OBL}\, \omega_e\, \gamma_b \,\Delta v_{0,\perp} m_e~,
\ee
which leads to the same scaling as in \eq{eperp}, if we take $\Delta v_{0,\perp}/c=\Delta p_{b,\rm OBL}/\gamma_b m_e c\sim \delta_{\rm OBL}$, as appropriate for the beam velocity dispersion at the end of the reactive oblique phase. In summary, apart from factors of order unity, the electric fields at the end of the kinetic oblique phase will saturate at a level similar to the reactive oblique stage (compare the two peaks of electric energy in \fig{timeevol}(a) at $\omega_e t\sim 600 $ and $\omega_e t\sim 1000 $). It follows that the kinetic oblique instability will increase the fraction of beam kinetic energy transferred to the plasma electrons only by a factor of order unity, as compared to the reactive oblique mode (see the orange line in \fig{timeevol}(a), for $\omega_e t\gtrsim1000$).

\begin{figure}[tbp]
\begin{center}
\includegraphics[width=0.5\textwidth]{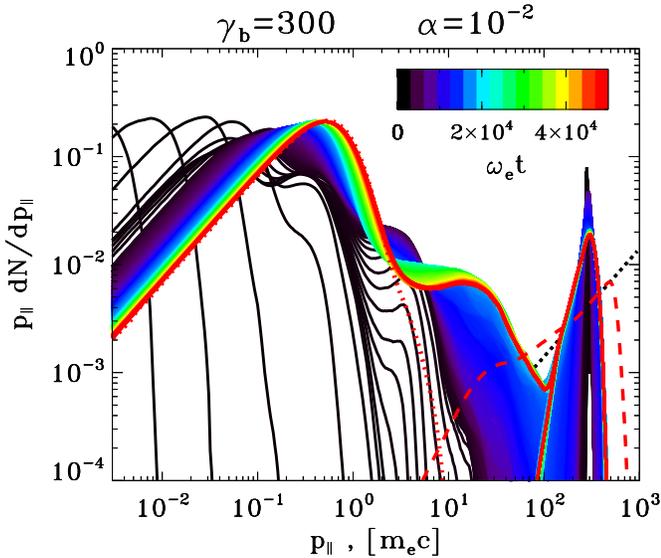}
\caption{Temporal evolution of the longitudinal momentum spectrum $p_{\parallel} dN/dp_\parallel$, for a beam-plasma system with $\gamma_b=300$ and $\alpha=\ex{2}$. The momentum is in units of  $m_e c$. The beam, which can be identified with the isolated peak at $p_\parallel\sim 300$, stops evolving after $\omega_e t\gtrsim1.5\times 10^4$, in agreement with \fig{timeevol}(f). The beam spectrum at late times does not relax to the so-called ``plateau'' distribution $dN/dp_\parallel\propto p_\parallel^0$, which is indicated as a black dotted line. We find that the beam approaches the plateau distribution only if we artificially inhibit the evolution of the beam transverse momentum, i.e., we force the beam relaxation to proceed in a quasi-1D configuration (red dashed line). As a result of the beam relaxation, the plasma develops a high-energy tail in the forward direction. For comparison, the plasma distribution in the backward direction (i.e., opposite to the beam) is shown as a dotted red line.}
\label{fig:momtime}
\end{center}
\end{figure}

\vspace{0.3in}
\subsubsection{The Longitudinal Relaxation Phase}\label{sec:relax}
After the saturation of the oblique instability, further evolution of the beam-plasma system proceeds via quasi-longitudinal modes, as shown in the 2D plot of the longitudinal electric field in \fig{timeevol}(g), as well as in the 3D pattern of \fig{3d}(c). Being quasi-longitudinal, these modes are insensitive to thermal spreads in the direction perpendicular to the beam, so they can grow even after the end of the kinetic oblique phase.

The quasi-longitudinal oscillations shown in \fig{timeevol}(g) are a characteristic signature of the {\it quasi-linear relaxation} of the beam \citep[see, e.g.,][]{grognard_75,lesch_87, sch_02,pavan_11}. In the quasi-linear relaxation, the beam generates longitudinal Langmuir waves (see the peak in $E_\parallel$  at $\ompt\sim10^4$ in \fig{timeevol}(e)), which scatter the beam particles and heat the background plasma (see the increase in the electron thermal energy shown by the orange line at $10^4\lesssim\ompt\lesssim1.5\times10^4$ in \fig{timeevol}(e)). Since $E_{\parallel}\gg E_{\perp}$ (compare the red and blue curves in \fig{timeevol}(e) at $\ompt\sim10^4$), the background electrons will be heated preferentially in the direction of motion of the beam (see the increase in $\Delta p_{e,\parallel}$ at $10^4\lesssim\ompt\lesssim1.5\times10^4$ in \fig{timeevol}(f)). For the same reason, the quasi-linear relaxation is accompanied by a substantial increase in the beam momentum spread along the direction of propagation (red solid line in  \fig{timeevol}(f), showing  the  growth of $\Delta p_{b,\parallel}$ at $5\times10^3\lesssim\ompt\lesssim1.5\times10^4$). The spread in the parallel beam momentum is associated to the formation of phase space holes, that result from the trapping of beam particles by the longitudinal electric oscillations \citep[e.g.,][]{oneil_71,thode_75}.

The quasi-linear relaxation occurs on a timescale much longer than the exponential oblique phase. Within the range of beam Lorentz factors and density contrasts probed by our simulations, we find that the characteristic relaxation time $\tau_{\rm R}$ is is at least two orders of magnitude longer than the exponential growth time of the oblique instability $\tau_{\rm OBL}=\omega_{\rm OBL}^{-1}$, in agreement with previous 1D simulations \citep{grognard_75,pavan_11}. This emphasizes the importance of evolving our PIC simulations to sufficiently long times to capture the physics of the quasi-linear relaxation (see \S\ref{sec:hot} for further details).

The quasi-linear modes broaden the beam  momentum spectrum in the longitudinal direction up to the point where $\Delta p_{b,\parallel}/\gamma_b m_e c\sim 0.2$ (see the red solid line in \fig{timeevol}(f), saturating at $\Delta p_{b,\parallel}/m_e c\sim 0.2\,\gamma_b\sim60$). This in agreement with the so-called Penrose's criterion, stating that the beam will be stable to electrostatic modes only when the longitudinal dispersion in momentum approaches the initial beam Lorentz factor \citep[e.g.,][]{buschauer_77}. From $\Delta p_{b,\parallel}/\gamma_b m_e c\sim 0.2$, it follows that at the end of the relaxation phase, a fraction $\sim 10\%$ of the beam energy has been transferred to the background electrons (see the orange line in \fig{timeevol}(e) at $\ompt\gtrsim2\times10^4$). In \S\ref{sec:cold} we demonstrate that, irrespective of the beam Lorentz factor or the beam-to-plasma density contrast, a generic by-product of the relaxation of cold ultra-relativistic dilute beams is the conversion of $\sim 10\%$ of their energy into plasma heating.\footnote{This is smaller than the heating efficiency of $\sim 30 \%$ reported by \citet{thode_75} using 1D simulations. In  Appendix \ref{sec:app}, we demonstrate that the transfer of beam energy to plasma electrons is indeed less efficient in 2D, as compared to the 1D case studied by \citet{thode_75}.}

The quasi-linear relaxation significantly affects the shape of the beam and plasma longitudinal momentum spectrum, as shown in \fig{momtime}. As a result of the quasi-longitudinal relaxation, the plasma distribution at late times ($\ompt\gtrsim2\times10^4$) develops a pronounced high-energy tail (at $3\lesssim p_\parr/m_e c \lesssim 10^2$), that bridges the main thermal peak of the plasma electrons (at $p_\parr/m_e c\sim 0.5$) with the beam particles (that populate the isolated high-energy bump at $p_\parr/m_e c\sim 300$ in \fig{momtime}). This high-energy component in the background electrons, which contains a significant amount of energy, is present only in the forward direction (i.e., along the beam propagation). In the backward direction, the spectrum of plasma electrons (red dotted line in \fig{momtime}, at $\ompt=5\times10^4$) is compatible with a Maxwellian.

During the quasi-linear relaxation,  the beam spectrum evolves from a quasi-monoenergetic distribution into a broad bump (from the black to the red curve at $p_\parr/m_e c\sim 300$ in \fig{momtime}). In agreement with \fig{timeevol}(f) (red solid line), most of the evolution occurs at $\ompt\lesssim 1.5\times 10^4$, whereas the beam spectrum at longer times is remarkably steady (we have followed the system up to $\ompt\sim1.5\times10^5$, finding no further signs of evolution). 

We point out that the beam momentum spectrum at late times does not approach the so-called ``plateau'' distribution $dN/dp_\parallel\propto p_\parallel^0$ (indicated as a black dotted line in \fig{momtime}), which is believed to be the ultimate outcome of the beam relaxation in 1D \citep[e.g.,][]{grognard_75,sch_02}. As opposed to earlier 1D claims, in our 2D and 3D simulations we find  that the beam longitudinal relaxation leads to a momentum spectrum that is harder than the plateau distribution, yet the beam-plasma system appears stable.\footnote{We have extensively checked that this result is numerically solid. We have confirmed our conclusions by using a larger number of computational particles per cell (up to 256), a larger 2D box (up to four times as large, in each direction), and a finer spatial resolution (up to $\comp=16$, instead of the usual value $\comp=8$).} In turn, the fact that the beam spectrum is harder than $dN/dp_\parallel\propto p_\parallel^0$ explains why the amount of beam energy transferred to the plasma is only $\sim 10\%$ (it should be $\sim 50\%$ for a plateau distribution extending up to $\gamma_b m_e c$, see \citealt{thode_75}). 

We argue  that the {\it transverse} dispersion in beam momentum, which could not be properly captured in previous 1D studies, prevents the {\it longitudinal} beam spectrum from relaxing to the plateau distribution (see Appendix \ref{sec:app2} for further details). Our claim is supported by the following experiment. At the end of the kinetic oblique phase ($\ompt\sim2000$), we artificially set the beam transverse dispersion to be $\Delta p_{b,\perp}/m_e c\ll1$ (for comparison, the self-consistent evolution in \fig{timeevol}(b) yields $\Delta p_{b,\perp}/m_e c\sim 10$ at the end of the kinetic oblique phase, see the solid blue line). Also, in the subsequent evolution, we inhibit any growth in the transverse beam momentum. In this setup, in which any transverse dispersion effects are artificially neglected, the beam relaxation should proceed as in 1D. So, it is not surprising that, in agreement with previous 1D studies, at late times the beam relaxes to the plateau distribution (red dashed line in \fig{momtime}). In other  words, we find that the relaxation to the plateau distribution is not a general result of the multi-dimensional evolution of ultra-relativistic beams, but it only occurs when the beam transverse dispersion stays sufficiently small, so the system is quasi-1D. 

As we further discuss in Appendix \ref{sec:app2}, the beam relaxation produces a plateau distribution only if $\Delta p_{b,\perp}/m_e c\ll1$, due to the following argument. Complete stabilization of the beam-plasma system is achieved when the longitudinal velocity spread of the ultra-relativistic beam reaches $\Delta v_{b,\parr}/c\sim1$ (more precisely, when the velocity spread is comparable to the beam speed). The spread in longitudinal velocity includes  contributions from both the longitudinal and the transverse momentum dispersions:
\be\label{eq:vspread}
\frac{\Delta v_{b,\parr}}{c}\sim \frac{\Delta p_{b,\parr}}{\gamma_b^3 m_e c} +\left(\frac{\Delta p_{b,\perp}}{\gamma_b m_e c}\right)^2~~,
\ee
where the second term on the right hand side is absent in the case of 1D relaxation. It follows that the transverse momentum spread can appreciably modify the relaxation process only if $\Delta p_{b,\perp}/m_e c\gtrsim\sqrt{\Delta p_{b,\parr}/\gamma_b m_e c}\sim 1 $, where we have used that $\Delta p_{b,\parr}/\gamma_b m_e c\sim 0.2$ at the end of the relaxation phase. Then, the fact that in the case studied in \fig{timeevol}(b) the transverse beam dispersion after the oblique phase is   $\Delta p_{b,\perp}/m_e c\sim 10$ explains why the beam relaxation cannot lead to a plateau distribution.  In Appendix \ref{sec:app2}, we provide further evidence that the transition to a plateau distribution requires $\Delta p_{b,\perp}/m_e c\ll1$.

\subsubsection{The Magnetic Field Growth}
In the previous subsections, we have primarily focused on the electrostatic character of the growing modes, which determines the coupling efficiency between the beam  energy and the plasma thermal energy. Here, we comment on the generation of magnetic fields associated with the evolution of the beam-plasma system.

As shown in \fig{timeevol}(a), the growth of the quasi-electrostatic oblique mode (both in the reactive and in the kinetic regime) is accompanied by a minor magnetic component. In \S\ref{sec:oblique}, we have estimated that the fraction of beam kinetic energy transferred to the magnetic fields at the end of the oblique phase is $\epsilon_{B,\perp}\sim \delta_{\rm OBL}^3$, apart from factors of order unity. Since $\delta_{\rm OBL}\ll1$ for ultra-relativistic dilute beams, the magnetic fields generated by the oblique instability are generally unimportant.

At the end of the relaxation phase, the beam and the plasma are highly anisotropic, with the longitudinal momentum spread much larger than the transverse one (see \fig{timeevol}(f) at $\ompt\sim1.5\times 10^4$). As a result, the system is prone to the Weibel instability \citep[e.g.,][]{weibel_59,yoon_87,silva_02}, which generates the transverse magnetic field pattern shown in \fig{timeevol}(h).\footnote{While the quasi-linear relaxation can also be captured with 1D simulations, the growth of Weibel modes necessarily requires multi-dimensional simulations.} As a result of the Weibel instability, the magnetic field energy increases (see the green line in \fig{timeevol}(e), at $10^4\lesssim\ompt\lesssim 2\times10^4$), and the beam and plasma anisotropy is reduced by increasing the transverse momentum spread (see the solid and dashed blue lines at $10^4\lesssim\ompt\lesssim 2\times10^4$ in \fig{timeevol}(f)).

The Weibel instability is predominantly magnetic, so it does not mediate any significant exchange of energy from the beam to the plasma electrons. Yet, it might be a promising source for the generation of magnetic fields, as discussed by \citet{sch12b}. However, the evolution of the magnetic filaments shown in \fig{timeevol}(h) can only be captured with large-scale 3D simulations \citep[e.g.,][]{bret_08,kong_09}. A detailed 3D investigation of the strength and scale of the magnetic fields resulting from ultra-relativistic dilute pair beams from TeV blazars will be presented elsewhere.


\begin{figure*}[tbp]
\begin{center}
\includegraphics[width=1\textwidth]{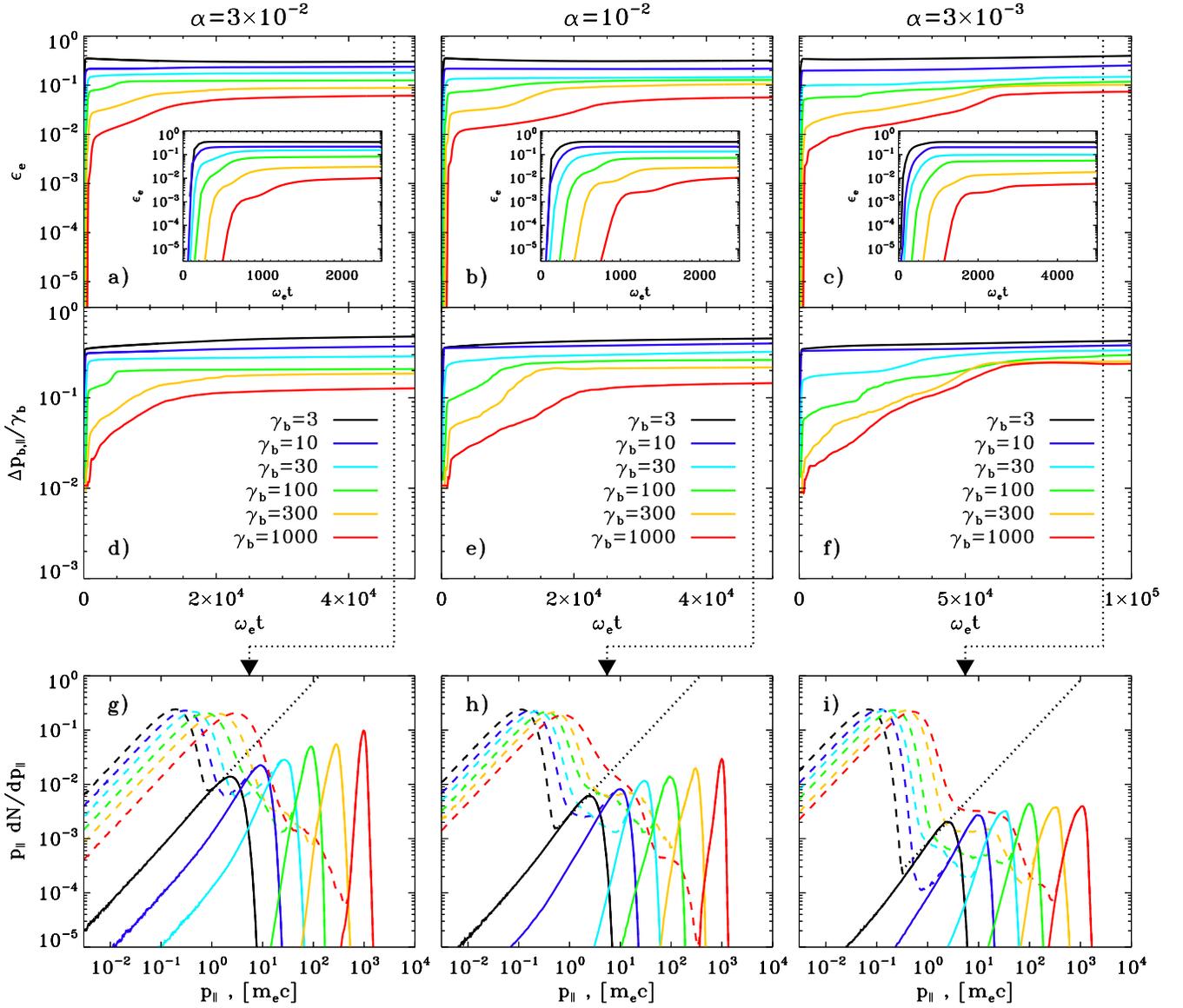}
\caption{Dependence of the beam-plasma evolution on the beam Lorentz factor (from $\gamma_b=3$ in black up to $\gamma_b=1000$ in red, in each panel; see the legend in panels (d)-(f)) and on the beam-to-plasma density contrast ($\alpha=3\times\ex{2}$ for the leftmost column, $\alpha=\ex{2}$ for the middle column, and $\alpha=3\times\ex{3}$ for the rightmost column). Panels (a)-(c):  fraction of the beam kinetic energy transferred to the plasma electrons (in the inset, a zoom-in on the earliest phases of evolution). Panels (d)-(f):  beam momentum dispersion in the longitudinal direction, normalized to the initial beam momentum  (i.e., $\Delta p_{b,\parr}/\gamma_b m_e c$). Panels (g)-(i): beam (solid) and total (dashed) momentum spectrum $p_{\parallel} dN/dp_\parallel$ in the longitudinal direction, at the time indicated with the dotted black lines in the upper rows. In panels (g)-(i), the slope $dN/dp_\parallel\propto p_\parr^0$ expected for the plateau distribution is shown as a dotted black line.}
\label{fig:ene}
\end{center}
\end{figure*}

\subsection{Dependence on the Beam Parameters}\label{sec:hot}
In this section, we discuss the dependence of the beam-plasma evolution on the beam parameters. In \S\ref{sec:cold}, we consider the case of cold beams, and we show that the quasi-longitudinal relaxation leads to a beam momentum spread along the direction of motion $\Delta p_{b,\parr}/\gamma_b m_e c\sim 0.2$, regardless of the beam Lorentz factor $\gamma_b$ or the beam-to-plasma density contrast $\alpha$. In turn, this implies that a fraction $\sim 10\%$ of the beam energy is converted into heat of the background plasma, irrespective of $\gamma_b$ or $\alpha$.

In \ref{sec:hothot}, we discuss the effect of the initial beam thermal spread on the efficiency of the beam-to-plasma energy transfer. We find that  if the initial dispersion in longitudinal momentum satisfies $\Delta p_{b0,\parr}/\gamma_b m_e c\gtrsim 0.2$ (as typically expected for blazar-induced beams), the fraction of beam energy deposited into the background plasma is much smaller than $\sim 10\%$. 

\subsubsection{Cold Beams}\label{sec:cold}
In \fig{ene} we show how the evolution of the beam-plasma system depends on the beam Lorentz factor (that we vary from $\gamma_b=3$ up to $\gamma_b=1000$) and on the beam-to-plasma density contrast (from $\alpha=3\times\ex{2}$ down to $\alpha=3\times\ex{3}$). Most of the previous studies have focused on moderately relativistic {\it electron} beams ($\gamma_b=3-6$) with $\alpha=\ex{1}$ \citep[e.g.,][]{gremillet_07,bret_08,kong_09}. Here, we extend our investigation to the case of ultra-relativistic dilute {\it electron-positron} beams, as appropriate for blazar-induced beams.

For each choice of $\gamma_b$ and $\alpha$, we follow the beam-plasma system from the oblique phase until the quasi-linear relaxation (typically, up to $\ompt\sim 10^5$).  We initialize a cold beam with thermal spread $k_{\rm B}T_b/m_e c^2\simeq\ex{4}$, so that the oblique instability initially proceeds in the reactive regime, for the range of $\gamma_b$ and $\alpha$ covered by our simulations. In the reactive phase, we confirm that the fastest growing mode has a wavevector oriented at $\sim 45^\circ$ to the beam direction of propagation. The growth rate is in excellent agreement with \eq{ommax}. As described in \S\ref{sec:oblique}, the exponential phase of the reactive oblique mode terminates due to self-heating of the beam in the transverse direction. At the end of the reactive phase, we find that the fractions of beam kinetic energy transferred to the plasma electrons, to the electric fields and to the magnetic fields scale respectively as $\epsilon_e\propto \delta_{\rm OBL}$, $\epsilon_{E,\perp}\sim \epsilon_{E,\parr}\propto \delta_{\rm OBL}$ and $\epsilon_{B,\perp}\propto \delta_{\rm OBL}^3$, as  in \S\ref{sec:oblique}.

The reactive oblique phase is followed by the kinetic oblique phase. We find that the characteristic wave pattern in the kinetic oblique phase (see \fig{timeevol}(d) in 2D and \fig{3d}(b) in 3D) appears in the evolution of all the beam-plasma systems that we present in \fig{ene}. In the temporal evolution of the fraction of beam energy converted into plasma heating, the kinetic oblique mode is responsible for the additional increase that is seen in the most relativistic cases ($\gamma_b\gtrsim100$) after the end of the reactive oblique phase  (see the insets in the top row of \fig{ene}). Regardless of $\gamma_b$ or $\alpha$, we find that the kinetic oblique phase deposits only a fraction $\sim\delta_{\rm OBL}$ of the beam kinetic energy into the background electrons, i.e., comparable to the reactive oblique phase (see \S\ref{sec:oblique}).

The long-term evolution of the system is controlled by the quasi-longitudinal relaxation, which operates on a timescale $\tau_{\rm R}\gtrsim 10^2\,\tau_{\rm OBL}$, where $\tau_{\rm OBL}=\omega_{\rm OBL}^{-1}$ is the characteristic {\it e}-folding time of the oblique mode. In \fig{ene}, the quasi-linear relaxation governs the growth in the plasma thermal energy (top row) and in the beam parallel momentum spread (middle row) occurring at $\omega_et\gtrsim5000$. In the regime $\gamma_b\gg1$, the quasi-linear relaxation terminates when the beam momentum spread in the longitudinal direction reaches $\Delta p_{b,\parr}/\gamma_b m_e c\sim0.2$, regardless of $\gamma_b$ or $\alpha$ (middle row in \fig{ene}). Correspondingly, the fraction of beam energy transferred to the plasma saturates at $\epsilon_e\sim 10\%$ (top row in \fig{ene}). 

Mildly relativistic beams with moderate density contrasts deviate from such simple scalings, for the following reason. The quasi-linear relaxation will not operate if the beam dispersion at the end of the oblique phase is already $\Delta p_{b,\parr}/\gamma_b m_e c\gtrsim0.2$. According to \eq{thresh}, the beam spread at the end of the oblique phase is $\Delta p_{b,\parr}\sim \Delta p_{b,\perp}\sim \delta_{\rm OBL}\gamma_b m_e c$, so that the quasi-linear relaxation will be suppressed if $\delta_{\rm OBL}\sim (\alpha/\gamma_b)^{1/3}\gtrsim 0.2$, i.e., for mildly relativistic beams with moderate $\alpha$.

A similar argument explains why a plateau distribution in the longitudinal momentum spectrum (bottom row in \fig{ene}) is established only for relatively small $\gamma_b$. As we have argued in \S\ref{sec:relax}, a  transverse spread  $\Delta p_{b,\perp}/m_e c\gtrsim 1$ prevents the beam relaxation to the plateau distribution (shown as a dotted black line in the bottom row of \fig{ene}). At the end of the oblique phase, $\Delta p_{b,\perp}/m_e c\sim \gamma_b\, \delta_{\rm OBL}$, so that the beam will relax to the plateau distribution only if $\gamma_b \delta_{\rm OBL}\sim (\gamma_b^2 \alpha)^{1/3}\ll1$. Clearly, this constraint is hardest to satisfy for highly relativistic beams, which explains why, at fixed $\alpha$, the momentum spectrum of  more relativistic beams shows stronger deviations from the plateau distribution.

\begin{figure}[tbp]
\begin{center}
\includegraphics[width=0.48\textwidth]{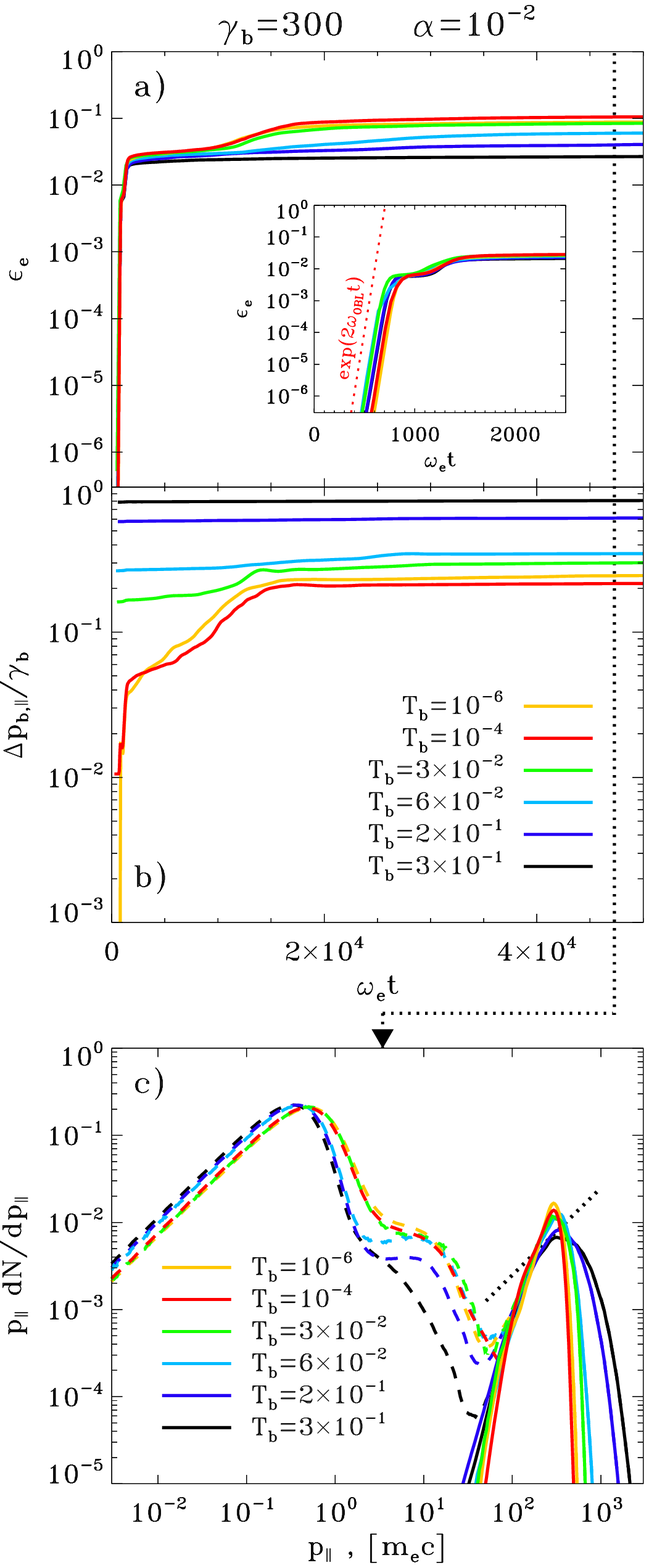}
\caption{Temporal evolution of a beam-plasma system with $\gamma_b=300$ and $\alpha=\ex{2}$, for different beam comoving temperatures at initialization (as shown by the legend in panel (b), where the beam temperature is in units of $m_e c^2/k_{\rm B}$. Panel (a): fraction of the beam kinetic energy deposited into the background electrons, with the inset showing the evolution at early times. Panel (b): temporal evolution of the beam longitudinal momentum spread, in units of $\gamma_b m_e c$. Panel (c): beam (solid) and total (dashed) momentum spectra in the longitudinal direction, at the time indicated in panels (a) and (b) with the vertical black dotted line. In panel (c), the dotted oblique line shows the slope expected for a plateau distribution $dN/dp_\parallel\propto p_\parr^0$.}
\label{fig:temp}
\end{center}
\end{figure}

\subsubsection{Hot Beams}\label{sec:hothot}
In the previous subsection, we have assumed that the beam is born with a negligible thermal spread. Here, we  discuss how the results presented above for cold beams will be modified by temperature effects. We describe separately the role of thermal spreads in the development of the oblique instability, and in the relaxation phase.

As we have anticipated in \S\ref{sec:oblique}, the oblique instability will proceed in the kinetic (rather than reactive) regime if the initial beam dispersion in the transverse direction is $\Delta v_{0,\perp}/c\gtrsim \delta_{\rm OBL}$.\footnote{The velocity dispersion $\Delta v_{0,\perp}$ can be recast as a comoving beam temperature $k_{\rm B}T_b/m_e c^2\sim(\gamma_b \Delta v_{0,\perp}/c)^2$, where we have assumed $k_{\rm B}T_b/m_e c^2\lesssim1$ (i.e., non-relativistic temperatures).} The growth rate in the kinetic regime is reported in \eq{kin}. The plasma thermal energy grows exponentially, until the self-generated electric fields increase the transverse dispersion in beam velocity beyond the initial value $\Delta v_{0,\perp}$. At this point, the exponential growth at the rate in \eq{kin} will necessarily terminate. From the Lorentz force applied to the beam particles, we find that this will happen when
\be
e E_{\perp}\sim \omega_k \gamma_b\, \Delta v_{0,\perp} m_e
\ee
which results in a fraction $\epsilon_{E,\perp}\sim \delta_k$ of the beam energy transferred to the electric fields at the end of the kinetic  phase ($\delta_k\equiv\omega_k/\omega_e$ is defined in \eq{kin}). Since the kinetic oblique mode is an electrostatic instability, the fraction of beam energy converted into heat will also be $\epsilon_e\sim \delta_k$. Moreover, due to the fact that $\omega_k\lesssim \omega_{\rm OBL}$ -- they are comparable  only if $\Delta v_{0,\perp}/c\sim \delta_{\rm OBL}$, i.e., at the boundary between reactive and kinetic regimes -- we expect the kinetic oblique mode to be less efficient in heating the plasma electrons, as compared to the reactive phase. For beams with $\gamma_b=3-10$ and $\alpha=\ex{3}-\ex{4}$, we have indeed verified with PIC simulations (not presented here) that the exponential growth of the kinetic oblique mode terminates at smaller $\epsilon_e$ for larger values of $\Delta v_{0,\perp}$, in good agreement with the expected scaling $\epsilon_e\propto \Delta v_{0,\perp}^{-2}$ (at fixed $\gamma_b$ and $\alpha$).\footnote{The condition $\Delta v_{0,\perp}/c\sim \sqrt{k_{\rm B} T_b /\gamma_b^2 m_e c^2}\gtrsim \delta_{\rm OBL}$, as required for the kinetic regime, together with the assumption of a quasi-monoenergetic beam (i.e., with comoving beam temperature $k_{\rm B} T_b/m_e c^2\lesssim 1$), constrains $\gamma_b\delta_{\rm OBL}\sim (\gamma_b^2\alpha)^{1/3}\lesssim1$, i.e., the kinetic regime can be best probed by low-$\gamma_b$ beams with $\alpha\ll1$.}

Regardless of the character of the oblique mode (reactive or kinetic), the long-term evolution of the beam-plasma system is controlled by the quasi-linear relaxation. In \fig{temp}, we show how the relaxation phase is affected by a finite beam temperature $T_b$. For the set of beam parameters employed in \fig{temp} ($\gamma_b=300$ and $\alpha=\ex{2}$), the oblique phase is expected to occur in the reactive regime, as long as the beam comoving temperature is non-relativistic. In fact, for all the choices of $T_b$ presented in \fig{temp} (in the plot, $T_b$ is in units of $m_e c^2/k_{\rm B}$), the early increase in the heating efficiency $\epsilon_e$ proceeds at the reactive rate $\omega_{\rm OBL}$ (compare the curves in the inset of \fig{temp}(a) with the dotted red line, that scales with the oblique growth rate). Also, the kinetic oblique instability, which is responsible for the further growth in $\epsilon_e$ at $\ompt\sim 1200$ (see the inset in \fig{temp}(a)), does not show any dependence on temperature, in the range $\ex{6}\lesssim k_{\rm B} T_b/m_e c^2\lesssim 0.3$ explored in \fig{temp}. 

The beam temperature has profound effects on the quasi-linear relaxation phase, for the beam parameters employed in \fig{temp}. For cold beams ($k_{\rm B} T_b/m_e c^2\lesssim \ex{4}$, yellow and red lines in \fig{temp}), the relaxation phase does not depend on the beam temperature. In agreement with the results presented in \S\ref{sec:cold}, the longitudinal spread in the beam momentum increases during the relaxation stage until $\Delta p_{b,\parr}/\gamma_b m_e c\sim 0.2$, as shown in \fig{temp}(b). This corresponds to a fraction $\epsilon_e\sim 10\%$ of the beam kinetic energy being converted into plasma heating (\fig{temp}(a)). Similar conclusions hold for moderate beam temperatures ($k_{\rm B}T_b/m_e c^2=3\times\ex{2}$, green line), whereas the quasi-linear relaxation is suppressed if the beam temperature at birth is such that the initial beam spread $\beamparz\gtrsim 0.2$ (cyan, blue and black lines in \fig{temp}(b)). In this case, the quasi-linear phase does not mediate any further increase in the heating efficiency $\epsilon_e$, beyond the early oblique phase (see the black line in \fig{temp}(a)).\footnote{Since the relaxation phase is quasi-longitudinal, the same conclusions hold in 1D. We have confirmed that, regardless of the nature of the fastest growing mode (oblique in 2D, longitudinal in 1D), the quasi-linear relaxation is suppressed if the beam spread at birth is such that $\beamparz\gtrsim 0.2$, both in 1D and in 2D.} In short, if the initial momentum spread is $\beamparz\gtrsim 0.2$,  the plasma heating efficiency $\epsilon_e$ stays fixed at the value $\epsilon_e\sim \delta_{\rm OBL}$ attained at the end of the reactive oblique phase -- or $\epsilon_e\sim \delta_{k}$, if the oblique instability proceeds in the kinetic regime.

The longitudinal momentum spectrum in \fig{temp}(c) clarifies why the quasi-linear relaxation is suppressed for hot beams. As we have discussed in \S\ref{sec:relax}, the relaxation is mediated by Langmuir waves excited by the beam. The beam particles that are slightly faster than the wave tend to transfer energy to the wave, and thus excite it, while those that are slightly slower than the wave tend to receive energy from it, and thus damp the wave. It follows that a mode is unstable if the number of beam particles moving slightly faster than the wave exceeds that of those moving slightly slower. More precisely, the instability will be stronger for a harder slope $d\log N/d\log p_\parr$ of the longitudinal momentum spectrum at momenta $\lesssim \gamma_b m_e c$ (this is the relevant momentum scale, since the phase velocity of the most unstable mode is slightly smaller than the beam speed). For a sharply peaked beam (i.e., with $k_{\rm B} T_b/m_e c^2\ll1$), the slope $d\log N/d\log p_\parr$ will be extremely hard, and the excitation of the Langmuir modes that mediate the relaxation process will be most efficient. As a result of the quasi-linear relaxation, the beam particles will be scattered down to lower energies, and when the beam momentum spread exceeds $\beampar\gtrsim0.2$ the slope of the momentum spectrum below the peak becomes too shallow (see the red  and yellow curves in \fig{temp}(c) at $p_\parr/m_e c\lesssim 300$), and the quasi-linear relaxation terminates. If the beam  temperature at birth is too hot (i.e., such that $\beamparz\gtrsim0.2$), the initial slope of the momentum spectrum at $p_\parr\lesssim \gamma_b m_e c$ is already too shallow to trigger efficient excitation of Langmuir waves, and the quasi-relaxation process is suppressed.

In \fig{temp}(c), we further support this argument by showing that, in the latest stages of evolution,  the shape of the beam momentum spectrum below $\gamma_b m_e c$ (in \fig{temp}(c), at $30\lesssim p_\parr/m_e c\lesssim300$), is nearly independent of the beam temperature at birth. For initially cold beams (yellow and red lines), the quasi-linear relaxation increases the beam momentum spread over time (see \fig{temp}(b)), until the beam spectrum relaxes to the broad bump shown in \fig{temp}(c). At this point, the beam is stable, although  it has not relaxed to the so-called plateau distribution (as we explain in \S\ref{sec:relax}, the relaxation to the plateau distribution requires $\Delta p_{b,\perp}/m_e c\ll1$, which is not satisfied here). Below $\gamma_b m_e c$, the {\it initial} momentum spectrum of hot beams resembles the {\it final} momentum distribution of initially cold beams. This explains why hot beams having  $\beampar\gtrsim0.2$ at birth will not experience the quasi-linear relaxation phase.

So far, we have discussed the effect of thermal spreads on the beam relaxation, assuming that the beam spectrum at birth is a drifting Maxwellian. However, the conclusions derived above hold for more complicated beam distributions. In particular, we have performed a set of PIC simulations, assuming that the beam spectrum at birth is a power law $dN/dp_\parr\propto p_\parr^{-2}$ for $p_{\parr}\geq p_{\rm min}\gg m_e c$. For blazar environments, such a beam spectrum is expected as a result of intense IC cooling off the CMB. 

If the beam power-law distribution has a negligible spread in the transverse direction (which might not be the case for blazar-induced beams, see \S\ref{sec:blazar}), then the oblique instability will proceed in the reactive regime. Apart from factors of order unity, we find that the exponential growth rate is $\sim (\alpha \,m_e c/p_{\rm min})^{1/3}$, similar to the case of mono-energetic beams. As we have discussed above, the effectiveness of the quasi-linear relaxation -- which ultimately determines whether the heating efficiency $\epsilon_e$ can reach $\sim 10\%$ -- is determined by the spread in the beam longitudinal spectrum {\it below the peak}, i.e., at $p_{\parallel}\lesssim p_{\rm min }$. If the beam spectrum has a sharp low-energy cutoff at $p_{\rm min}$, then the quasi-linear relaxation does operate, and the beam deposits $\sim 10\%$ of its energy into the background electrons. However, if the low-energy end of the beam distribution is broader, the quasi-linear relaxation will be inhibited, and the amount of beam energy transferred to the plasma will be much smaller, in agreement with the results  in \fig{temp}. We find that, if the beam longitudinal spectrum below $p_{\rm min}$ can be modeled as a power law $dN/dp_\parr\propto p_{\parr}^s$, the quasi-linear relaxation is suppressed if $s\lesssim 3$, with the case $s=0$ corresponding to the plateau distribution.


\subsection{Implications for Blazar-Induced Beams}\label{sec:blazar}
We now analyze the implications of our findings for the evolution of blazar-induced beams in the IGM. As we have discussed in \S\ref{sec:setup}, the Lorentz factors and density contrasts of blazar-driven beams  ($\gamma_b\sim10^6-10^7$ and $\alpha\sim\ex{18}-\ex{15}$) cannot be directly studied with PIC simulations. However, by performing a number of experiments with a broad range of $\gamma_b\gg1$ and $\alpha\ll1$, we have been able to assess how the relaxation of ultra-relativistic dilute beams depends on the beam Lorentz factor and the beam-to-plasma density ratio. Our results can then be confidently extrapolated to the extreme parameters of blazar-induced beams. 

We have demonstrated that the oblique instability, which governs the earliest stages of evolution of ultra-relativistic dilute beams, proceeds in the kinetic regime if the initial velocity dispersion in the direction transverse to the beam is $\Delta v_{0,\perp}/c\gtrsim \delta_{\rm OBL}$, where $\delta_{\rm OBL}\equiv\omega_{\rm OBL}/\omega_e\sim (\alpha/\gamma_b)^{1/3}$ is the growth rate of the reactive oblique mode in units of the plasma frequency $\omega_e=\sqrt{4\pi e^2 n_e/m_e}\simeq20\, (n_e/10^{-7}\rm cm^{-3})^{1/2}$ rad$\,$s$^{-1}$ of the IGM electrons. Since the beam pairs are born with a mildly relativistic thermal spread in the center of mass of the photon-photon interaction (see \S\ref{sec:model}), the transverse velocity spread in the IGM frame will be $\Delta v_{0,\perp}/c\sim 1/\gamma_b$. If follows that the oblique instability will proceed in the kinetic regime if $\gamma_b \delta_{\rm OBL}\lesssim 1$, which is  marginally satisfied for blazar-induced beams ($\gamma_b \delta_{\rm OBL}\sim0.01-1$). The oblique instability will grow at the kinetic rate $\omega_k\sim \omega_e (c/\Delta v_{0,\perp})^2 \alpha/\gamma_b $ in \eq{kin}, which for $\Delta v_{0,\perp}/c\sim 1/\gamma_b$, as appropriate for blazar-induced beams, reduces to $\omega_k\sim \gamma_b\, \alpha\, \omega_e$. For the parameters of blazar-induced beams, this is a factor of 
\be\label{eq:oIC}
\frac{\omega_k}{c/d_{\rm IC}}\simeq 10^5 \left(\frac{\alpha}{\ex{16}}\right)\left(\!\frac{n_e}{10^{-7}\unit{cm^{-3}}}\!\right)^{1/2}
\ee
larger than the IC cooling rate $c/d_{\rm IC}$, where $d_{\rm IC}\simeq 100\,(\gamma_b/10^7)^{-1}\unit{kpc}$ is the IC cooling length computed in \S\ref{sec:model}. In agreement with \citet{brod12a} and \citet{sch12b}, we find that the kinetic oblique instability has ample time to grow, before the beam loses energy to IC emission. Furthermore, the typical lifetime of blazar activity of $\sim 10^7\unit{years}$ $\gg d_{\rm IC}/c$ is sufficient for the instability to operate.\footnote{However, jet variability may be fast enough to compete with the instability growth time. Our analysis focuses on the steady or long-term average TeV emission.} 

Due to self-heating of the beam in the transverse direction (see \S\ref{sec:time} and \ref{sec:hothot}), the exponential phase of the kinetic oblique instability terminates when only a minor fraction of the beam energy has been transferred to the IGM electrons. As we have argued in \S\ref{sec:hothot}, the heating efficiency of blazar-induced beams at the end of the kinetic oblique phase is only 
\be\label{eq:heat}
\epsilon_e\sim \delta_k\equiv\frac{\omega_k}{\omega_e}\simeq \ex{9}\left(\frac{\gamma_b}{10^7}\right)\left(\frac{\alpha}{\ex{16}}\right)~~.
\ee
It follows that, even though the oblique instability can grow faster than the IC cooling time, its efficiency in heating the plasma electrons is extremely poor.

As we have emphasized in \S\ref{sec:relax}, a larger amount of beam energy (up to $\sim 10\%$) can be deposited into the plasma electrons by the quasi-linear relaxation phase. We find that the quasi-linear relaxation occurs on a timescale much longer than the growth time of the oblique instability, $\tau_{\rm R}\gtrsim 10^2 c/\omega_k$. Yet, since the oblique growth rate is much faster than the IC cooling rate, see \eq{oIC}, the quasi-linear relaxation should have enough time to operate before the beam energy is lost to IC emission. Here, we are conservatively neglecting the possibility that $\tau_{\rm R}$ might be much longer than $\sim10^2\, c/\omega_k$ for more extreme beam parameters, as a result of nonlinear plasma processes that reduce the strength of the electric fields available for the quasi-linear relaxation \citep[see][]{sch12b}.

In \S\ref{sec:hothot}, we have demonstrated that the quasi-linear relaxation can occur only if the beam momentum spectrum along the longitudinal direction is sufficiently narrow. More precisely, we find that if the beam distribution peaks at $\gamma_b m_e c$ (in the case of a power-law tail, this would be the low-energy cutoff), the quasi-linear relaxation can operate only if the parallel momentum spread {\it below the peak} satisfies $\Delta p_{b0,\parr}/\gamma_b m_e c\lesssim 0.2$. This constraint is hard to fulfill by blazar-induced beams. Since the pair creation cross section peaks slightly above the threshold energy, the pairs are born moderately warm, with a comoving temperature $k_B T_b/m_e c^2\sim 0.5$. This corresponds to a longitudinal momentum spread at birth (measured in the IGM frame) of $\Delta p_{b0,\parr}/\gamma_b m_e c\sim 1$, which is already prohibitive for the development of the quasi-linear relaxation. Moreover, the momentum dispersion of blazar-induced beams might be  even larger, since the spectrum of both the EBL and the blazar TeV emission are usually modeled as broad power laws \citep{miniati_13}.

In summary, a solid assessment of the shape of the beam momentum distribution below the peak is essential to predict the amount of beam energy deposited into the IGM. If the beam spectrum at birth were to have $\Delta p_{b0,\parr}/\gamma_b m_e c\lesssim 0.2$  (which is unlikely to be the case, but see \citealt{sch12b}), then the quasi-linear relaxation would transfer $\epsilon_e\sim 10\%$ of the beam energy to the IGM. Even in this optimistic case, we remark that the heating efficiency would reach at most $\sim 10\%$, so $\sim 90\%$ of the energy would still remain in the beam. At the end of the quasi-linear relaxation phase, the beam spectrum will   be harder than the so-called plateau distribution, since the transverse dispersion in beam velocity does not meet the requirement $\Delta v_{0,\perp}/c\ll1/\gamma_b$ for relaxation to the plateau spectrum.  In the more realistic case $\Delta p_{b0,\parr}/\gamma_b m_e c\gtrsim 0.2$, the quasi-linear relaxation will be inhibited, resulting in a lower efficiency of IGM heating -- with a firm lower limit being the heating fraction at the end of the oblique phase, see \eq{heat}.


\subsubsection{Comparison with Earlier Studies}
We now compare our numerical work with earlier analytical studies of the relaxation of blazar-induced beams in the IGM. As we have emphasized in \S\ref{sec:setup}, our PIC simulations are the first to address the evolution of dilute {\it ultra-relativistic}  {\it electron-positron} beams, as appropriate for blazar-induced beams in the IGM. Most of the previous PIC studies \citep[e.g.,][]{dieckmann_06,gremillet_07,bret_08,kong_09} have focused on {\it mildly relativistic} {\it electron} beams.

In agreement with \citet{brod12a} and \citet{sch12a}, we find that the fastest growing instability for blazar-induced beams propagating through the IGM is the oblique mode. If the pair beam were to be initially cold, the oblique instability would evolve at the reactive rate in \eq{ommax},  with wavevector oriented at $\sim 45^\circ$ relative to the beam (see \citealt{sch12a}). However, the initial transverse spread in beam momentum is large enough such that the oblique mode evolves at the kinetic rate in \eq{kin}, as argued by \citet{brod12a} and \citet{miniati_13}. We remark that the transverse beam spread does  affect the growth of the {\it oblique} mode, but it will not impact the evolution of {\it longitudinal}  waves, as found by \citet{sch13}. Yet, since the early evolution of blazar-induced beams is controlled by the oblique (rather than longitudinal) mode, transverse thermal effects are indeed important for the beam-plasma interaction at early times.

The evolution of  blazar-induced beams at late times is governed by non-linear plasma processes, which are extremely hard to capture with analytical tools. \citet{sch12b} and \citet{miniati_13} attempted to describe the non-linear relaxation of blazar-induced beams, reaching opposite conclusions regarding the ultimate fate of the beam energy. \citet{sch12a} assumed that the beam momentum distribution can be modeled as a delta function  (i.e., they approximated the beam as mono-energetic and uni-directional), and they found that the relaxation phase occurs much faster than the IC cooling losses. From this, they argued that more than $50\%$ of the beam energy can be transferred to the IGM plasma.  \citet{miniati_13} reached the opposite conclusion, when accounting for the effect of the finite transverse momentum spread of blazar-induced beams. They found that the beam energy is radiated by IC off the CMB well before the relaxation phase.

With our PIC simulations, we find that the relaxation phase occurs on a much longer timescale than the exponential oblique growth, at least by two orders of magnitude. However, it might be delayed even more for more extreme beam parameters, as suggested by both \citet{sch12b} and \citet{miniati_13}. Even under the conservative assumption that the relaxation phase is faster than the IC cooling time, this does not imply that all of the beam energy is ultimately deposited into the IGM plasma. In short, the relaxation process being {\it faster} than the IC losses does not guarantee that it will also be {\it efficient} in heating the IGM electrons. Under the unrealistic assumption that blazar-driven beams are born with a small longitudinal momentum spread $\beamparz\lesssim 0.2$, we find that only $\sim 10\%$ (rather than $\sim 50\%-100\%$, as assumed by \citet{brod12a}) of the beam energy  is transferred to the plasma. For the realistic  spread  $\beamparz\sim 1$ of blazar-induced beams, the coupling will be much less efficient, with the heating efficiency as low as $\epsilon_e\simeq \ex{9} (\gamma_b/10^7) (\alpha/\ex{16})$.

\section{Summary and Discussion}\label{sec:astro}
The interaction of TeV photons from distant blazars with the extragalactic background light produces ultra-relativistic electron-positron pairs. The resulting pair beam is unstable to the excitation of beam-plasma instabilities in the unmagnetized intergalactic medium (IGM). The ultimate fate of the beam energy is uncertain, and it is hard to capture with analytical tools. By means of 2D and 3D PIC simulations, we have investigated the linear and non-linear evolution of ultra-relativistic dilute electron-positron beams. We have performed dedicated experiments with a broad range of beam Lorentz factors $\gamma_b=3-1000$ and beam-to-plasma density contrasts $\alpha=\ex{3}-\ex{1}$, so that our results can be extrapolated to the extreme parameters of blazar-induced beams ($\gamma_b=10^6-10^7$ and $\alpha=\ex{18}-\ex{15}$). 

We find that the earliest stages of evolution of  ultra-relativistic dilute beams are governed by the oblique instability. For cold beams, the oblique mode proceeds in the so-called reactive regime, where all the beam particles are interacting with the unstable waves, and the coupling between the beam and the plasma is most efficient. In this case, the instability grows at the reactive rate $\omega_{\rm OBL}=\delta_{\rm OBL}\omega_e$, where $\delta_{\rm OBL}\sim (\alpha/\gamma_b)^{1/3}$ and $\omega_e=\sqrt{4\pi e^2 n_e/m_e}\simeq20\,(n_e/10^{-7}\unit{cm^{-3}})^{1/2} \unit{rad\, s^{-1}}$ is the plasma frequency of the IGM electrons. However, blazar-induced beams are not cold. The initial spread in transverse velocity is $\Delta v_{0,\perp}\sim 1/\gamma_b\gtrsim \delta_{\rm OBL}$, so that the oblique mode proceeds in the kinetic (rather than reactive) regime. Here, the spectrum of unstable modes is broader, with only a few beam particles being in resonance with each mode. The growth rate is $\omega_k=\delta_{k}\,\omega_e\sim \omega_e (c/\Delta v_{0,\perp})^2 \gamma_b\, \alpha$, which is smaller than the corresponding rate $ \omega_{\rm OBL}$ of cold beams, yet large enough such that the kinetic oblique instability has ample time to grow, before the beam loses energy to IC emission by scattering off the CMB. 

On the other hand, the oblique instability growing {\it faster} than the IC losses does not guarantee that it will also be {\it efficient} in heating the IGM electrons. Due to self-heating of the beam in the transverse direction, we find that the exponential phase of the kinetic oblique instability terminates when only a minor fraction of the beam energy has been transferred to the IGM plasma. At the end of the kinetic oblique phase, the heating efficiency of IGM electrons  is extremely poor, reaching $\epsilon_e\sim \delta_k\simeq \ex{9} (\gamma_b/10^7) (\alpha/\ex{16})$. 

Additional transfer of energy from the beam to the plasma occurs at later times (with a delay of two or more orders of magnitude, relative to the oblique growth time), and it is mediated by the quasi-linear relaxation process. Here, the beam generates longitudinal electrostatic waves, which scatter the beam particles -- thus broadening the beam momentum spectrum in the longitudinal direction -- and heat the IGM electrons. The quasi-linear relaxation can operate only if the beam momentum spectrum along the longitudinal direction is sufficiently narrow. More precisely, we find that if the beam distribution peaks at $\gamma_b m_e c$, the quasi-linear relaxation requires   the parallel momentum spread below the peak to be  $\Delta p_{b0,\parr}/\gamma_b m_e c\lesssim 0.2$. In this case, a fraction $\epsilon_e\sim 10\%$ of the beam energy is transferred to the plasma (rather than $\sim 50\%- 100\%$, as assumed by \citealt{brod12a}).

 The constraint $\Delta p_{b0,\parr}/\gamma_b m_e c\lesssim 0.2$ on the longitudinal dispersion at birth  is hard to fulfill by blazar-induced beams. Since the pair creation cross section peaks slightly above the threshold energy, the pairs are born moderately warm, with a comoving temperature $k_B T_b/m_e c^2\sim 0.5$. This corresponds to a longitudinal momentum spread at birth (measured in the IGM frame) of $\Delta p_{b0,\parr}/\gamma_b m_e c\sim 1$, which is already prohibitive for the development of the quasi-linear relaxation. Moreover, the momentum dispersion of blazar-induced beams might be  even larger, since the spectrum of both the EBL and the blazar TeV emission are usually modeled as broad power laws  \citep{miniati_13}. For $\Delta p_{b0,\parr}/\gamma_b m_e c\sim 1$, the quasi-linear relaxation will be suppressed, which results in a much lower efficiency of IGM heating -- with a firm lower limit being the heating fraction at the end of the oblique phase $\epsilon_e\sim \delta_k\ll1$.

A the end of the relaxation phase,  the beam and plasma distributions are highly anisotropic, with the longitudinal momentum spread much larger than the transverse one. As a result, the system is prone to the Weibel instability \citep[e.g.,][]{weibel_59,yoon_87,silva_02}, which relaxes the beam and plasma anisotropy by generating  transverse magnetic fields. The Weibel instability is predominantly magnetic, so it does not mediate any further exchange of energy from the beam to the plasma electrons. Yet, it might be a promising source for the generation of magnetic fields in the IGM, as discussed by \citet{sch12b}. A multi-dimensional PIC investigation of the strength and scale of the magnetic fields resulting from blazar-induced beams will be presented elsewhere.

Our results have important implications for the ultimate fate of the energy of blazar-induced beams. Since at most $\sim 10\%$ of the beam energy is deposited into the IGM plasma, most of the energy ($\gtrsim 90\%$) is still available for IC interactions with the CMB. Therefore, the {\it Fermi}  non-detection of the IC scattered GeV emission around TeV blazars can be reliably used to probe the strength of the EBL and of the IGM magnetic fields. In particular, the   lower bounds on the IGM field strength derived by various authors \citep[e.g.,][]{neronov_10,tavecchio_10} are still valid, despite the fast growth of beam-plasma instabilities in the IGM. However, when computing the expected IC signature, one should take into account that beam-plasma instabilities tend to broaden the beam distribution function. As a result,  the reprocessed GeV emission will be spread over a wider frequency range (and consequently, with smaller flux), as compared to the case of mono-energetic beams.



The fraction of beam energy deposited into the IGM might have important cosmological implications, as argued by \citet{brod12b} and \citet{brod12c}. By assuming that {\it all} of the beam energy is transferred to the IGM electrons by beam-plasma instabilities, they showed that blazar heating could dominate over photo-heating in the low-redshift evolution of the IGM, by almost one order of magnitude. If the heating efficiency of blazar-induced beams is $\sim 10\%$, rather than $\sim 100\%$ as they assumed,  blazar heating could still contribute as much as photo-heating to the thermal history of the IGM. However, we expect the heating efficiency of blazar-induced beams to be $\epsilon_e\ll1$, for the following reasons:
\begin{itemize}
\item As we have argued above, blazar-induced beams are born with a significant longitudinal spread in momentum, $\beamparz\sim 1$. In this case, the quasi-linear relaxation process, which mediates efficient transfer of the beam energy to the IGM electrons up to $\epsilon_e\sim 10\%$, will be suppressed, and the heating fraction remains $\epsilon_e\sim \delta_k\ll1$.
\item Even if the relaxation process were to be operating, the beam relaxation timescale might be much longer than the IC loss time, as argued by \citet{miniati_13} (but see \citet{sch12b}, for opposite conclusions).
\item Density inhomogeneities associated with cosmic structure induce loss of resonance between the beam particles and the excited plasma oscillations, strongly inhibiting the growth of the unstable modes \citep{miniati_13}.
\item A large-scale IGM field might spread a mono-energetic uni-directional beam in the transverse and longitudinal directions, thus inhibiting the efficiency of the beam-plasma interaction.\footnote{For simplicity, we neglect the fact that the presence of a large-scale field might change the nature of the fastest growing instabilities, as compared to the unmagnetized case explored in this work.} In the oblique phase, a sufficiently strong magnetic field might trigger the transition to the kinetic regime, if during the growth time of the reactive instability ($\sim \omega_{\rm OBL}^{-1}$) it deflects the beam velocity sideways by more than the threshold $\sim \delta_{\rm OBL} c$ between the reactive and kinetic regimes. This happens if $\omega_{\rm B}/\omega_e\gtrsim \delta_{\rm OBL}^2$, where $\omega_{\rm B}= eB/\gamma_b m_e c$ is the Larmor frequency of the beam particles in the IGM fields. The limit on the IGM field strength is
\be
\!\!\!\!\!\!B\gtrsim  5\!\times\! 10^{-15}\!\left(\!\frac{\gamma_b}{10^7}\!\right)^{1/3}\!\!  \left(\!\frac{\alpha}{\ex{16}}\!\right)^{2/3} \!\! \left(\!\frac{n_e}{10^{-7}\unit{cm^{-3}}}\!\right)^{1/2}\!\!\!\!\unit{G}
\ee
which is generally realized for IGM fields \citep[e.g.,][]{neronov_09}. 

As regards to the quasi-linear relaxation, IGM fields are expected to change the evolution of the quasi-linear process if during the characteristic quasi-linear growth time $\tau_{\rm R}$ the field deflects the beam such that the longitudinal dispersion in momentum reaches $\beampar\gtrsim 0.2$. This can be recast as $\omega_{\rm B}\tau_{\rm R}\gtrsim 1$. A weaker constraint can be derived by using that the quasi-linear relaxation plays a role only if the IC cooling time is $d_{\rm IC}/c\gtrsim \tau_{\rm R}$. By setting $\omega_{\rm B}\, d_{\rm IC}/c\gtrsim 1$, we obtain a limit on the field strength that inhibits the quasi-linear relaxation  
\be
B\gtrsim  5\times 10^{-14} \left(\frac{\gamma_b}{10^7}\right)^{2}\unit{G}
\ee
which might be satisfied by IGM fields \citep[e.g.,][]{neronov_09}. 

\end{itemize}
For these reasons, we conclude that blazar-induced beams are not likely to play a major role in the thermal history of the IGM.



\acknowledgements
We thank A. Bret for insightful comments. L.S. is supported by NASA through Einstein
Postdoctoral Fellowship grant number PF1-120090 awarded by the Chandra
X-ray Center, which is operated by the Smithsonian Astrophysical
Observatory for NASA under contract NAS8-03060. The simulations were
performed on XSEDE resources under
contract No. TG-AST120010, and on NASA High-End Computing (HEC) resources through the NASA Advanced Supercomputing (NAS) Division at Ames Research Center.

\begin{figure}[tbp]
\begin{center}
\includegraphics[width=0.5\textwidth]{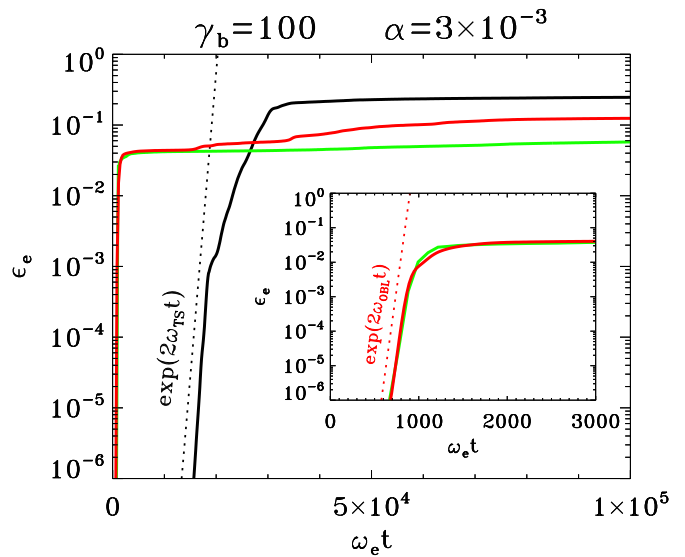}
\caption{Comparison between our 2D simulation (red line) with two 1D simulations (black and green lines), such that the computational box is oriented at two different angles relative to the beam velocity. The beam has $\gamma_b=100$ and $\alpha=3\times\ex{3}$. For the black line, the 1D computational box is oriented along the beam, i.e., the simulation only selects the unstable modes whose wavevector is parallel to the beam. For the green line, the 1D domain forms an angle $\theta_{\rm box}=45^\circ$ with the beam velocity, so that only the oblique modes can be captured by the 1D simulation. The inset shows the evolution of the heating efficiency $\epsilon_e$ at early times.}
\label{fig:1dtilt}
\end{center}
\end{figure}

\begin{figure}[tbp]
\begin{center}
\includegraphics[width=0.48\textwidth]{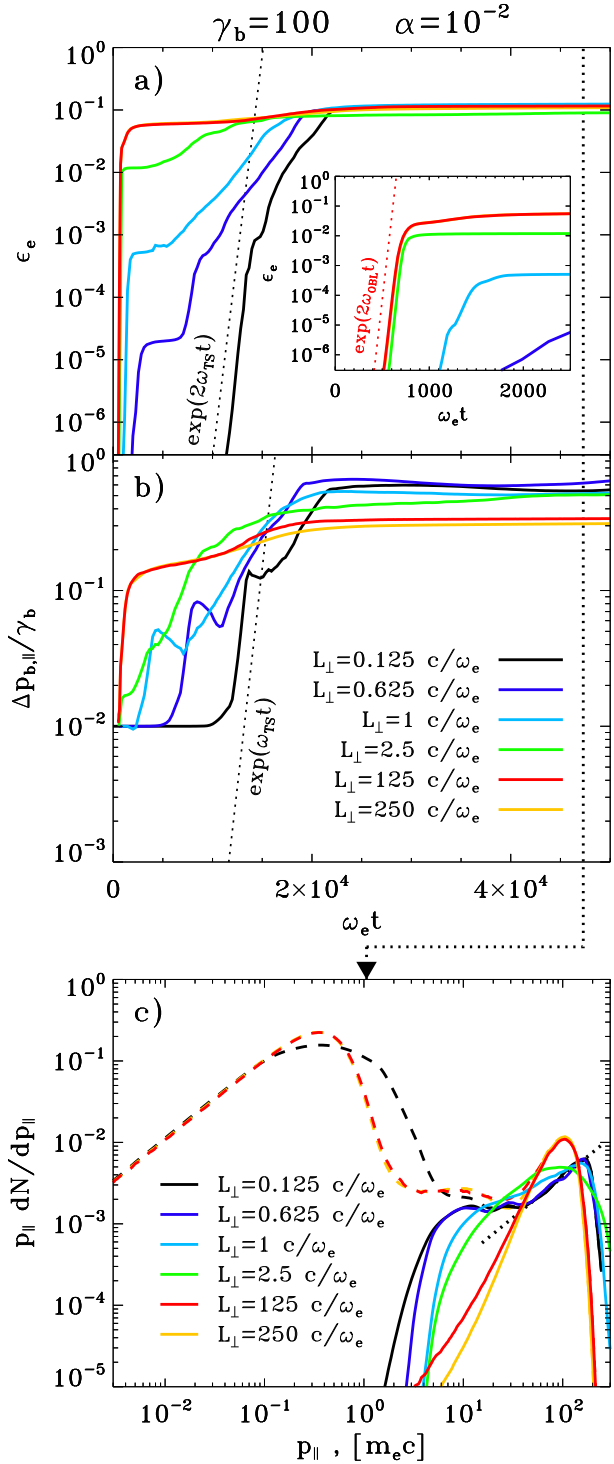}
\caption{Temporal evolution of the beam-plasma interaction, as a function of the box size $L_\perp$ in the direction transverse to the beam (instead, $L_\parr=125\comp$ along the beam in all cases). The beam has $\gamma_b=100$ and $\alpha=\ex{2}$. Panel (a): temporal evolution of the heating efficiency $\epsilon_e$, with the inset showing the evolution at early times. Panel (b): temporal evolution of the beam longitudinal momentum spread, in units of $\gamma_b m_e c$. Panel (c): the solid lines show the beam momentum spectrum in the longitudinal direction, at the time indicated in panels (a) and (b) with the vertical black dotted line. For three selected cases ($L_\perp=0.125\comp$, $125\comp$ and $250\comp$), we also plot the total (beam plus plasma) momentum spectra with dashed lines. In panel (c), the dotted oblique line shows the slope expected for a plateau distribution $dN/dp_\parallel\propto p_\parr^0$.}
\label{fig:1d2d}
\end{center}
\end{figure}

\appendix
\section{A. 1D Simulations of Relativistic Dilute Beams}\label{sec:app}
In the main body of the paper, we have presented our results on the relaxation of ultra-relativistic dilute beams, by employing 2D and 3D simulations. Here, we discuss how the physics of the beam-plasma evolution differs, when performing 1D simulations. 

In \fig{1dtilt}, we compare our 2D results (red line) with two selected 1D simulations (black and green curves), that differ in the orientation of the beam relative to the simulation box. For the black line, the beam is aligned with the simulation domain, whereas the two directions form an angle of $\theta_{\rm box}=45^\circ$ for the green line. Since the fastest growing oblique mode for ultra-relativistic dilute cold beams is oriented at $\sim 45^\circ$ relative to the beam propagation (see \S\ref{sec:oblique}), the 1D box at an angle $\theta_{\rm box}=45^\circ$ with respect to the beam  should correctly capture the evolution of the oblique mode. This is confirmed by the inset in \fig{1dtilt}, which shows that the exponential growth in the heating efficiency  $\epsilon_e$ proceeds at the expected rate $\omega_{\rm OBL}$ of \eq{ommax} both in 2D (red line) and in the 1D simulation with $\theta_{\rm box}=45^\circ$ (green).

On the other hand,  the 1D simulation with $\theta_{\rm box}=45^\circ$ cannot correctly capture the relaxation phase, which is mediated by quasi-longitudinal modes. In fact, the heating efficiency $\epsilon_e$ in the 1D box with $\theta_{\rm box}=45^\circ$ does not significantly change after the end of the oblique phase. In contrast, as a result of the quasi-longitudinal relaxation,  in 2D the heating fraction $\epsilon_e$ increases at $10^4\lesssim\ompt\lesssim 5\times 10^4$ up to the saturation value $\epsilon_e\sim 10\%$. 

The quasi-linear relaxation, being driven by longitudinal modes, can be described in 1D with a simulation box oriented along the beam (black line in \fig{1dtilt}). For a 1D box with  $\theta_{\rm box}=0^\circ$ (black line), the quasi-linear relaxation  controls the beam evolution at $2\times10^4\lesssim\ompt\lesssim 3\times 10^4$. At late times, the heating efficiency saturates at $\epsilon_e\sim 20\%$ (in agreement with \citealt{thode_75}), which is twice as large as compared to the analogous 2D case. As anticipated in \S\ref{sec:relax}, this is related to the role of transverse spreads in the relaxation of ultra-relativistic beams. In multi-dimensions, the longitudinal velocity spread $\Delta v_{b,\parr}$ required to terminate the relaxation phase can be achieved either by decelerating the beam in the longitudinal direction (with a fractional energy loss $\Delta p_{b,\parr}/\gamma_b m_e c$), or by deflecting the beam sideways (which gives a transverse spread $\Delta p_{b,\perp}$, but no significant energy loss). The contribution of the two terms is presented in \eq{vspread}. A 1D box with $\theta_{\rm box}=0^\circ$ can only capture beam-aligned modes, which cannot change the transverse spread $\Delta p_{b,\perp}$, so that the second term in \eq{vspread} does not contribute. It follows that the same $\Delta v_{b,\parr}$ will be attained in 1D by a larger $\Delta p_{b,\parr}/\gamma_b m_e c$ (and so, higher $\epsilon_e$), relative to its 2D counterpart. This explains why in 1D (for beam-aligned boxes) the heating fraction $\epsilon_e$ is a factor of a few larger than in 2D (compare black and red lines in \fig{1dtilt}).

In summary, beam-aligned 1D simulations tend to overestimate the fraction of beam energy deposited into the background electrons by the relaxation phase. Most importantly, they cannot properly model the early evolution of the beam-plasma system, which is mediated by oblique modes. Rather, the exponential phase in 1D simulations with $\theta_{\rm box}=0^\circ$ will necessarily proceed at the two-stream growth rate $\omega_{\rm TS}=\gamma_b^{2/3}\omega_{\rm OBL}$, which is indicated as a dotted black line in \fig{1dtilt}. In short, the multi-dimensional physics of the beam-plasma evolution cannot be properly captured by 1D simulations.

The difference between 1D and 2D simulations is also presented in \fig{1d2d}, where we discuss the dependence of our results on the transverse size of the computational domain, from $L_{\perp}=0.125\comp$ (1D simulation) to our standard choice $L_{\perp}=125\comp$. We also confirm that our 2D results are the same when doubling the box size in the transverse direction (compare the red lines for $L_{\perp}=125\comp$ with the yellow lines for $L_{\perp}=250\comp$). 

For 1D boxes aligned with the beam, the two-stream instability governs the exponential growth of $\epsilon_e$ at early times (compare the black solid and dotted lines in \fig{1d2d}(a) and (b)). The oblique mode can operate only if the transverse size of the box is $L_\perp\gtrsim 2.5\comp$, as shown by the fact that the green line in the inset of \fig{1d2d}(a) grows at the oblique rate $\omega_{\rm OBL}$ indicated by the dotted red line. For a box with $L_\perp=0.625 \comp$ (blue line), the oblique phase mediates the growth of $\epsilon_e$ at early times ($\ompt\lesssim3000$), yet at a rate smaller than $\omega_{\rm OBL}$, whereas the exponential stage of the two-stream mode  emerges at later times ($\ompt\sim7000$) with the expected rate $\omega_{\rm TS}$ (see the blue line in \fig{1d2d}(a) and (b)).

In agreement with \fig{1dtilt}, we find that the quasi-linear relaxation in 2D proceeds in a similar way as in 1D, apart from the fact that the dispersion in longitudinal momentum at late times is smaller for larger box widths, as shown in \fig{1d2d}(b). In turn, this is related to the shape of the beam momentum distribution at the end of the quasi-linear relaxation phase. As shown in  \fig{1d2d}(c), the longitudinal momentum spectrum in 1D simulations relaxes to the plateau distribution $dN/dp_\parallel\propto p_\parr^0$ (indicated as a dotted black line in \fig{1d2d}(c)), whereas in 2D the beam spectrum below the peak stays harder than the plateau distribution. As we have argued in \ref{sec:relax} (see Appendix \ref{sec:app2} for further details), the difference between our 1D and 2D results is ultimately related to the transverse spread in beam momentum, which is larger in 2D than in 1D.

\begin{figure}[tbp]
\begin{center}
\includegraphics[width=0.5\textwidth]{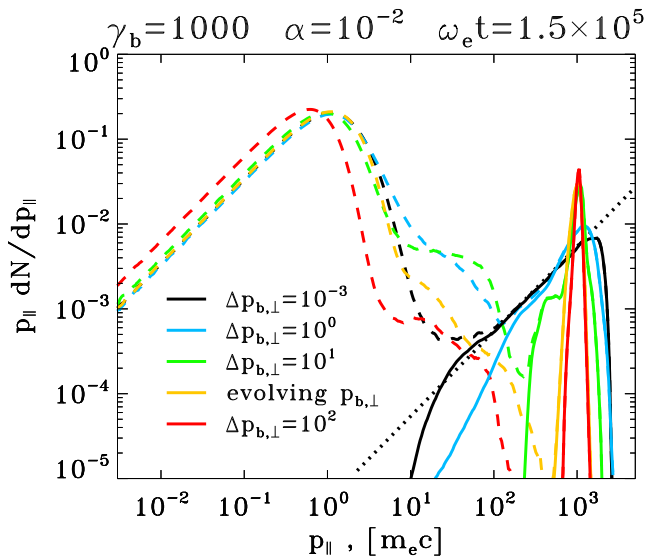}
\caption{Longitudinal momentum spectrum at late times ($\ompt=1.5\times 10^5$), for a beam-plasma system with $\gamma_b=1000$ and $\alpha=\ex{2}$. We perform the following experiment: after the end of the oblique phase, we artificially reset the transverse spread in beam momentum to the values indicated in the legend (there, $\Delta p_{b,\perp}$ is in units of $m_e c$), and we inhibit by hand its evolution. For the case indicated with the yellow line, we allow the beam to evolve without constraints (i.e., this would correspond to the self-consistent evolution of the beam-plasma system), which results in $\Delta p_{b,\perp}/m_e c\sim 20$ at late times. It is apparent that  relaxation to the plateau distribution (indicated as a dotted black line) requires $\Delta p_{b,\perp}/m_e c\ll1$.}
\label{fig:relax}
\end{center}
\end{figure}

\section{B.  Relaxation to the Plateau Distribution}\label{sec:app2}
In \S\ref{sec:relax}, we have argued, based on \eq{vspread}, that the longitudinal velocity spread $\Delta v_{b,\parr}$ required to terminate the relaxation process can be sourced not only by a longitudinal spread in momentum $\Delta p_{b,\parr}$, but also by a transverse spread $\Delta p_{b,\perp}$. In 1D the latter is absent, whereas in 2D it is the interplay of the two that  determines the shape of the beam momentum spectrum at late times. In particular, this explains why in 2D the beam does not necessarily relax to the plateau distribution, which instead is a general outcome of the beam-plasma evolution in 1D configurations.

\fig{relax} clarifies the role of the transverse momentum spread for the relaxation to the plateau distribution. Using 2D simulations, we perform the following experiment. Right after the end of the oblique phase, we artificially reset the transverse momentum spread $\Delta p_{b,\perp}$ to the values indicated by the legend. Also, in the course of the subsequent evolution, we inhibit its evolution by hand. We find that, if we artificially set $\Delta p_{b,\perp}/m_e c\ll1$, the beam relaxation proceeds as in 1D, and the longitudinal momentum spectrum at late times relaxes to the plateau distribution (compare the black solid and dotted lines in \fig{relax}). In contrast, if $\Delta p_{b,\perp}/m_e c\gtrsim 1$ the quasi-linear relaxation does spread the beam momentum in the longitudinal direction, but not enough to approach the plateau distribution. We have verified that the threshold $\Delta p_{b,\perp}/m_e c\sim1$ for  relaxation to the plateau dsitribution holds irrespective of the beam Lorentz factor or the beam-to-plasma density contrast.

As a result of the growth of the oblique mode, the transverse dispersion of cold beams at the end of the oblique phase approaches $\Delta p_{b,\perp}/m_e c\sim \gamma_b \delta_{\rm OBL}$. For the beam parameters employed in \fig{relax} ($\gamma_b=1000$ and $\alpha=\ex{2}$), this would give $\Delta p_{b,\perp}/m_e c\sim 20$. This explains why the beam spectrum plotted as a  yellow solid curve in \fig{relax}, which corresponds to the self-consistent evolution of the beam-plasma system (i.e., the beam transverse dispersion is not constrained by hand), does not approach the plateau distribution at late times. 

\bibliography{IGM}

\begin{thebibliography}{49}
\expandafter\ifx\csname natexlab\endcsname\relax\def\natexlab#1{#1}\fi

\bibitem[{{Abdo} {et~al.}(2010){Abdo}, {Ackermann}, \& {Ajello}}]{abdo_10}
{Abdo}, A.~A., {Ackermann}, M., \& {Ajello}, M. e.~a. 2010, \apj, 723, 1082

\bibitem[{{Aharonian} {et~al.}(2006){Aharonian}, {Akhperjanian}, \&
  {Bazer-Bachi}}]{aharonian_06}
{Aharonian}, F., {Akhperjanian}, A.~G., \& {Bazer-Bachi}, A.~R. e.~a. 2006,
  \nat, 440, 1018

\bibitem[{{Aharonian}(2001)}]{aharonian_01}
{Aharonian}, F.~A. 2001, in International Cosmic Ray Conference, Vol.~27,
  International Cosmic Ray Conference, I250

\bibitem[{{Bret} {et~al.}(2005){Bret}, {Firpo}, \& {Deutsch}}]{bret_05}
{Bret}, A., {Firpo}, M.-C., \& {Deutsch}, C. 2005, \pre, 72, 016403

\bibitem[{{Bret} {et~al.}(2010{\natexlab{a}}){Bret}, {Gremillet}, \&
  {B{\'e}nisti}}]{bret_10b}
{Bret}, A., {Gremillet}, L., \& {B{\'e}nisti}, D. 2010{\natexlab{a}}, \pre, 81,
  036402

\bibitem[{{Bret} {et~al.}(2008){Bret}, {Gremillet}, {B{\'e}nisti}, \&
  {Lefebvre}}]{bret_08}
{Bret}, A., {Gremillet}, L., {B{\'e}nisti}, D., \& {Lefebvre}, E. 2008,
  Physical Review Letters, 100, 205008

\bibitem[{{Bret} {et~al.}(2010{\natexlab{b}}){Bret}, {Gremillet}, \&
  {Dieckmann}}]{bret_10}
{Bret}, A., {Gremillet}, L., \& {Dieckmann}, M.~E. 2010{\natexlab{b}}, Physics
  of Plasmas, 17, 120501

\bibitem[{{Bre{\v i}zman} \& {Ryutov}(1971)}]{breizman_ryutov_71}
{Bre{\v i}zman}, B.~N. \& {Ryutov}, D.~D. 1971, Soviet Journal of Experimental
  and Theoretical Physics, 33, 220

\bibitem[{{Broderick} {et~al.}(2012){Broderick}, {Chang}, \&
  {Pfrommer}}]{brod12a}
{Broderick}, A.~E., {Chang}, P., \& {Pfrommer}, C. 2012, \apj, 752, 22

\bibitem[{{Broderick} {et~al.}(2013){Broderick}, {Pfrommer}, {Puchwein}, \&
  {Chang}}]{brod13}
{Broderick}, A.~E., {Pfrommer}, C., {Puchwein}, E., \& {Chang}, P. 2013,
  ArXiv:1308.0340

\bibitem[{{Buneman}(1958)}]{buneman_58}
{Buneman}, O. 1958, Physical Review Letters, 1, 8

\bibitem[{{Buneman}(1993)}]{buneman_93}
---. 1993, {in ``Computer Space Plasma Physics'', Terra Scientific, Tokyo, 67}

\bibitem[{{Buschauer} \& {Benford}(1977)}]{buschauer_77}
{Buschauer}, R. \& {Benford}, G. 1977, \mnras, 179, 99

\bibitem[{{Chang} {et~al.}(2012){Chang}, {Broderick}, \& {Pfrommer}}]{brod12b}
{Chang}, P., {Broderick}, A.~E., \& {Pfrommer}, C. 2012, \apj, 752, 23

\bibitem[{{Dermer} {et~al.}(2011){Dermer}, {Cavadini}, {Razzaque}, {Finke},
  {Chiang}, \& {Lott}}]{dermer_11}
{Dermer}, C.~D., {Cavadini}, M., {Razzaque}, S., {Finke}, J.~D., {Chiang}, J.,
  \& {Lott}, B. 2011, \apjl, 733, L21

\bibitem[{{Dieckmann} {et~al.}(2006{\natexlab{a}}){Dieckmann}, {Frederiksen},
  {Bret}, \& {Shukla}}]{dieckmann_06b}
{Dieckmann}, M.~E., {Frederiksen}, J.~T., {Bret}, A., \& {Shukla}, P.~K.
  2006{\natexlab{a}}, Physics of Plasmas, 13, 112110

\bibitem[{{Dieckmann} {et~al.}(2006{\natexlab{b}}){Dieckmann}, {Shukla}, \&
  {Drury}}]{dieckmann_06}
{Dieckmann}, M.~E., {Shukla}, P.~K., \& {Drury}, L.~O.~C. 2006{\natexlab{b}},
  \mnras, 367, 1072

\bibitem[{{Dolag} {et~al.}(2011){Dolag}, {Kachelriess}, {Ostapchenko}, \&
  {Tom{\`a}s}}]{dolag_11}
{Dolag}, K., {Kachelriess}, M., {Ostapchenko}, S., \& {Tom{\`a}s}, R. 2011,
  \apjl, 727, L4

\bibitem[{{Fainberg} {et~al.}(1970){Fainberg}, {Shapiro}, \&
  {Shevchenko}}]{fainberg_70}
{Fainberg}, Y.~B., {Shapiro}, V.~D., \& {Shevchenko}, V.~I. 1970, Soviet
  Journal of Experimental and Theoretical Physics Letters, 30, 528

\bibitem[{{Ghisellini} {et~al.}(2010){Ghisellini}, {Tavecchio}, {Foschini},
  {Ghirlanda}, {Maraschi}, \& {Celotti}}]{ghisellini_10}
{Ghisellini}, G., {Tavecchio}, F., {Foschini}, L., {Ghirlanda}, G., {Maraschi},
  L., \& {Celotti}, A. 2010, \mnras, 402, 497

\bibitem[{{Gremillet} {et~al.}(2007){Gremillet}, {B{\'e}nisti}, {Lefebvre}, \&
  {Bret}}]{gremillet_07}
{Gremillet}, L., {B{\'e}nisti}, D., {Lefebvre}, E., \& {Bret}, A. 2007, Physics
  of Plasmas, 14, 040704

\bibitem[{{Grognard}(1975)}]{grognard_75}
{Grognard}, R.~J.-M. 1975, Australian Journal of Physics, 28, 731

\bibitem[{{Kong} {et~al.}(2009){Kong}, {Park}, {Ren}, {Sheng}, \&
  {Tonge}}]{kong_09}
{Kong}, X., {Park}, J., {Ren}, C., {Sheng}, Z.~M., \& {Tonge}, J.~W. 2009,
  Physics of Plasmas, 16, 032107

\bibitem[{{Lemoine} \& {Pelletier}(2010)}]{pelletier_10}
{Lemoine}, M. \& {Pelletier}, G. 2010, \mnras, 402, 321

\bibitem[{{Lesch} \& {Schlickeiser}(1987)}]{lesch_87}
{Lesch}, H. \& {Schlickeiser}, R. 1987, \aap, 179, 93

\bibitem[{{Miniati} \& {Elyiv}(2013)}]{miniati_13}
{Miniati}, F. \& {Elyiv}, A. 2013, \apj, 770, 54

\bibitem[{{Nakar} {et~al.}(2011){Nakar}, {Bret}, \&
  {Milosavljevi{\'c}}}]{nakar_11}
{Nakar}, E., {Bret}, A., \& {Milosavljevi{\'c}}, M. 2011, \apj, 738, 93

\bibitem[{{Neronov} \& {Semikoz}(2009)}]{neronov_09}
{Neronov}, A. \& {Semikoz}, D.~V. 2009, \prd, 80, 123012

\bibitem[{{Neronov} \& {Vovk}(2010)}]{neronov_10}
{Neronov}, A. \& {Vovk}, I. 2010, Science, 328, 73

\bibitem[{{O'Neil} {et~al.}(1971){O'Neil}, {Winfrey}, \& {Malmberg}}]{oneil_71}
{O'Neil}, T.~M., {Winfrey}, J.~H., \& {Malmberg}, J.~H. 1971, Physics of
  Fluids, 14, 1204

\bibitem[{{Pavan} {et~al.}(2011){Pavan}, {Yoon}, \& {Umeda}}]{pavan_11}
{Pavan}, J., {Yoon}, P.~H., \& {Umeda}, T. 2011, Physics of Plasmas, 18, 042307

\bibitem[{{Pfrommer} {et~al.}(2012){Pfrommer}, {Chang}, \&
  {Broderick}}]{brod12c}
{Pfrommer}, C., {Chang}, P., \& {Broderick}, A.~E. 2012, \apj, 752, 24

\bibitem[{{Puchwein} {et~al.}(2012){Puchwein}, {Pfrommer}, {Springel},
  {Broderick}, \& {Chang}}]{brod12d}
{Puchwein}, E., {Pfrommer}, C., {Springel}, V., {Broderick}, A.~E., \& {Chang},
  P. 2012, \mnras, 423, 149

\bibitem[{{Schlickeiser} {et~al.}(2012{\natexlab{a}}){Schlickeiser}, {Elyiv},
  {Ibscher}, \& {Miniati}}]{sch12a}
{Schlickeiser}, R., {Elyiv}, A., {Ibscher}, D., \& {Miniati}, F.
  2012{\natexlab{a}}, \apj, 758, 101

\bibitem[{{Schlickeiser} {et~al.}(2012{\natexlab{b}}){Schlickeiser}, {Ibscher},
  \& {Supsar}}]{sch12b}
{Schlickeiser}, R., {Ibscher}, D., \& {Supsar}, M. 2012{\natexlab{b}}, \apj,
  758, 102

\bibitem[{{Schlickeiser} {et~al.}(2013){Schlickeiser}, {Krakau}, \&
  {Supsar}}]{sch13}
{Schlickeiser}, R., {Krakau}, S., \& {Supsar}, M. 2013, \apj, 777, 49

\bibitem[{{Schlickeiser} {et~al.}(2002){Schlickeiser}, {Vainio},
  {B{\"o}ttcher}, {Lerche}, {Pohl}, \& {Schuster}}]{sch_02}
{Schlickeiser}, R., {Vainio}, R., {B{\"o}ttcher}, M., {Lerche}, I., {Pohl}, M.,
  \& {Schuster}, C. 2002, \aap, 393, 69

\bibitem[{{Silva} {et~al.}(2002){Silva}, {Fonseca}, {Tonge}, {Mori}, \&
  {Dawson}}]{silva_02}
{Silva}, L.~O., {Fonseca}, R.~A., {Tonge}, J.~W., {Mori}, W.~B., \& {Dawson},
  J.~M. 2002, Physics of Plasmas, 9, 2458

\bibitem[{{Sironi} \& {Spitkovsky}(2011)}]{sironi_spitkovsky_11a}
{Sironi}, L. \& {Spitkovsky}, A. 2011, \apj, 726, 75

\bibitem[{{Sironi} {et~al.}(2013){Sironi}, {Spitkovsky}, \&
  {Arons}}]{sironi_13}
{Sironi}, L., {Spitkovsky}, A., \& {Arons}, J. 2013, \apj, 771, 54

\bibitem[{{Spitkovsky}(2005)}]{spitkovsky_05}
{Spitkovsky}, A. 2005, in AIP Conf. Ser., Vol. 801, Astrophysical Sources of
  High Energy Particles and Radiation, ed. {T.~Bulik, B.~Rudak, \&
  G.~Madejski}, 345

\bibitem[{{Takahashi} {et~al.}(2012){Takahashi}, {Mori}, {Ichiki}, \&
  {Inoue}}]{takahashi_12}
{Takahashi}, K., {Mori}, M., {Ichiki}, K., \& {Inoue}, S. 2012, \apjl, 744, L7

\bibitem[{{Tavecchio} {et~al.}(2010){Tavecchio}, {Ghisellini}, {Foschini},
  {Bonnoli}, {Ghirlanda}, \& {Coppi}}]{tavecchio_10}
{Tavecchio}, F., {Ghisellini}, G., {Foschini}, L., {Bonnoli}, G., {Ghirlanda},
  G., \& {Coppi}, P. 2010, \mnras, 406, L70

\bibitem[{{Taylor} {et~al.}(2011){Taylor}, {Vovk}, \& {Neronov}}]{taylor_11}
{Taylor}, A.~M., {Vovk}, I., \& {Neronov}, A. 2011, \aap, 529, A144

\bibitem[{{Thode}(1976)}]{thode_76}
{Thode}, L.~E. 1976, Physics of Fluids, 19, 305

\bibitem[{{Thode} \& {Sudan}(1975)}]{thode_75}
{Thode}, L.~E. \& {Sudan}, R.~N. 1975, Physics of Fluids, 18, 1552

\bibitem[{{Vovk} {et~al.}(2012){Vovk}, {Taylor}, {Semikoz}, \&
  {Neronov}}]{vovk_12}
{Vovk}, I., {Taylor}, A.~M., {Semikoz}, D., \& {Neronov}, A. 2012, \apjl, 747,
  L14

\bibitem[{{Weibel}(1959)}]{weibel_59}
{Weibel}, E.~S. 1959, Physical Review Letters, 2, 83

\bibitem[{{Yoon} \& {Davidson}(1987)}]{yoon_87}
{Yoon}, P.~H. \& {Davidson}, R.~C. 1987, \pra, 35, 2718

\end{thebibliography}

\end{document}